\documentclass[aps,prx,preprint,onecolumn,amsmath,amssymb,superscriptaddress,nofootinbib]{revtex4-1}

\immediate\write18{cp $(kpsewhich apsrev4-1.bst) apsrev4-1-allauthors.bst}

\usepackage[english]{babel}
\usepackage{amsthm}
\usepackage{booktabs}
\usepackage{graphicx}
\usepackage{tikz}
\usepackage{algorithm}
\usepackage{comment}
\usepackage{url}
\usepackage{qcircuit}
\usepackage{algpseudocode}
\usepackage{xcolor}
\usepackage[colorlinks=true,linkcolor=blue,citecolor=blue,urlcolor=blue]{hyperref}

\graphicspath{{figures/}}

\newcommand{\Hank}{\mathcal{H}}
\newcommand{\Toep}{\mathcal{C}}
\newcommand{\Om}{\Omega}
\newcommand{\PiOm}{\Pi_{\Omega}}
\newcommand{\Cc}{\mathbb{C}}
\newcommand{\Rr}{\mathbb{R}}
\newcommand{\Tmax}{T_{\mathrm{max}}}

\newcommand{\Nshots}{N_{\mathrm{shots}}}
\newcommand{\eps}{\epsilon}
\newcommand{\ket}[1]{\left|#1\right\rangle}
\newcommand{\bra}[1]{\left\langle#1\right|}
\newcommand{\braket}[2]{\left\langle#1\middle|#2\right\rangle}
\newcommand{\norm}[1]{\left\lVert#1\right\rVert}

\begin{document}

\title{Molecular energies from sparse quantum prolate diagonalization on a trapped-ion quantum processor}

\author{Davide Castaldo}
\email{dcastaldo@ethz.ch}
\affiliation{Department of Chemistry and Applied Biosciences, ETH Zurich, Vladimir-Prelog-Weg 2, 8093 Zurich, Switzerland}

\author{Marvin Kronenberger}
\affiliation{Department of Chemistry and Applied Biosciences, ETH Zurich, Vladimir-Prelog-Weg 2, 8093 Zurich, Switzerland}

\author{Marek Miller}
\affiliation{Quantum for Life Centre, Department of Mathematical Sciences, University of Copenhagen,
Universitetsparken 5, 2100 Copenhagen \O{}, Denmark}

\author{S\'el\`ene Forget}
\affiliation{
NNF Quantum Computing Programme, Niels Bohr Institute, University of Copenhagen, Blegdamsvej 17, 2100 Copenhagen \O{}, Denmark}
\affiliation{Linderstr{\o}m-Lang Centre for Protein Science, University of Copenhagen, Ole Maaløes Vej 5, 2200 Copenhagen N, Denmark}

\author{Gemma C. Solomon}
\affiliation{
NNF Quantum Computing Programme, Niels Bohr Institute, University of Copenhagen, Blegdamsvej 17, 2100 Copenhagen \O{}, Denmark}
\affiliation{Nano-Science Center and Department of Chemistry, University of Copenhagen, Universitetsparken 5, 2100 Copenhagen \O{}, Denmark}

\author{Kresten Lindorff-Larsen}
\affiliation{Linderstr{\o}m-Lang Centre for Protein Science, University of Copenhagen, Ole Maaløes Vej 5, 2200 Copenhagen N, Denmark}

\author{Matthias Christandl}
\email{christandl@math.ku.dk}
\affiliation{Quantum for Life Centre, Department of Mathematical Sciences, University of Copenhagen,
Universitetsparken 5, 2100 Copenhagen \O{}, Denmark}

\author{Markus Reiher}
\email{mreiher@ethz.ch}
\affiliation{Department of Chemistry and Applied Biosciences, ETH Zurich, Vladimir-Prelog-Weg 2, 8093 Zurich, Switzerland}

\date{November 17, 2026}

\begin{abstract}
\small
We introduce sparse quantum prolate diagonalization (sparse QPD), which combines the original formulation of QPD [Phys.\ Rev.\ Res., 8(3), 033177, (2026)] with compressed sensing to exploit Hamiltonian spectrum sparsity for molecular systems. We
show that sparse sampling reduces the sample complexity of QPD from $\mathcal{O}(N_s)$ to $\mathcal{O}(\log N_s)$,
where $N_s$ is the number of sampled points of the autocorrelation function (i.e.\ distinct circuits),
without increasing circuit depth at chemical accuracy.
With a two-orbital active-space-embedded Hamiltonian for an interacting arginine--tryptophan residue pair in water as an example,
we show that on the Quantinuum Helios trapped-ion quantum computer,
sparse QPD resolves the Hamiltonian energies to chemical accuracy with circuits of up to 448 two-qubit gates. For a suitably large quantum processor, we demonstrate numerically in the weakly correlated regime that the method extends to larger systems.
\end{abstract}

\maketitle

\section{Introduction}
\label{sec:intro}

An open question in quantum chemistry is whether ab initio calculations on quantum hardware can enhance or even accelerate high-throughput computational workflows~\cite{fauseweh2024quantum, castaldo2026utility, babbush2026grand}.
Despite recent advances in quantum error correction,
the significant overhead of fault-tolerant execution still dominates the runtime of quantum algorithms~\cite{gratsea2025achieving, dalzell2025quantum}.
Moreover, such overhead scales with circuit volume, rendering error correction increasingly cumbersome as the number of logical operations grows\,\cite{aharonov2025importance, bharti2026fault}. 

The practical application of quantum computers to chemistry~\cite{von2021quantum,capone2024vision} requires algorithms that minimize physical-qubit and logical-operation costs. Textbook quantum phase estimation (QPE)~\cite{nielsen2000quantum}, which can deliver accurate electronic energies with known error bounds, is not the optimal solution in this regard, as its connectivity requirements and circuit depth lead to substantial resource overheads~\cite{reiher2017elucidating,delgado2022simulating}. Despite significant reductions enabled by advances in error correction and algorithmic techniques~\cite{low2025fast,caesura2025faster,low2026denser}, achieving scalability competitive with the best classical quantum chemistry heuristics~\cite{raghavachari1989fifth} remains an open challenge. 

This has spurred extensive research efforts to improve on the phase estimation algorithm by using single-ancilla quantum circuits~\cite{ding2023even, ding2023simultaneous, li2023adaptive, stroeks2022spectral, gunther2026phase, dutkiewicz2022heisenberg, ding2024quantum}.
Single-ancilla phase estimation algorithms are hybrid methods in which the quantum computer samples the system's autocorrelation function, while classical post-processing extracts eigenvalues of the unitary matrix. In this context, the total runtime on the quantum computer is characterized by the number of independent circuits to be executed and by the maximal evolution time $T_{max}$ required to achieve $\epsilon$-accurate energy estimate. In Ref.\,\cite{stroschein2026ground}, it was shown that  quantum prolate diagonalization (QPD) can achieve chemically accurate energy estimates for both weakly and strongly correlated systems (i.e., those with an error within 1.6 mHa) using evolution times more than one order of magnitude shorter than those required by standard QPE, while requiring at least three orders of magnitude fewer shots per sample than other quantum subspace methods\,\cite{klymko2022real,cohn2021quantum,cortes2022quantum,lee2024sampling,oliveira2025quantum,shen2025estimating}.

In this paper, we continue to reduce the overall cost of QPD for a target error $\epsilon$, numerically demonstrating that sparse sampling combined with matrix completion~\cite{Cai2025} can reduce the number of autocorrelation function samples, $N_s$, from $\mathcal{O}(N_s)$ to $\mathcal{O}(\log N_s)$. We will refer to this variant as \emph{sparse QPD}. The idea of using compressed sensing techniques~\cite{candes2006near, candes2010matrix, candes2010power, candes2014towards} in a single-ancilla phase estimation algorithm was first proposed for ground-state phase estimation in Ref.\,\cite{yi2024quantum}, and subsequently extended to the estimation of multiple eigenvalues in Ref.\,\cite{castaldo2025heisenberg}. Here, we show its application to a quantum subspace method reporting up to an order of magnitude reduction of the total algorithmic cost without increasing the circuit depth.

Our second contribution in this work is to evaluate sparse QPD under realistic levels of shot noise, Trotter error, and hardware noise, including experiments on the Quantinuum Helios trapped-ion device (Helios-1) and its emulator (Helios-1E)~\cite{ransford2026,decross2025,mayer2023}. We build on previous experimental demonstrations of QPE for chemistry~\cite{o2016scalable, blunt2023statistical, yamamoto2024demonstrating, yamamoto2026quantum, nutzel2026ground, kanno2025tensor, sorathia2025quantum}. Using standard first-order Trotterized Hamiltonian simulation, which is well suited to small molecular systems, we achieve energy errors below 1~kcal/mol on hardware for circuits containing up to 448 two-qubit gates (1221 gates in total).

This paper is organized as follows. In Sec.\,\ref{sec:theory}, we show how to incorporate the signal recovery algorithm of Ref.\,\cite{Cai2025} into the original QPD algorithm. We then investigate the performance of sparse QPD relative to standard QPD. Our numerical analysis in Sec.\,\ref{sec:sparse_sampling_qpd} is supported by a Cramér–Rao lower bound derived in App.\,\ref{app:crlb}, following the approach of Refs.\,\cite{dutkiewicz2022heisenberg, kimura2026fisher}. Next, we present the results of our experiment on the Helios machine and characterize the performance of the method in the presence of hardware noise (Sec.\,\ref{quantinuum_results}). Finally, we conclude with some remarks and discuss directions for future research.

\section{Theory}
\label{sec:theory}

The following section outlines the theory behind sparse QPD: Sec.~\ref{sec:pfd} recalls quantum prolate diagonalization to obtain energy estimates from equidistant samples of the autocorrelation function $C(t_k)$. Sec.~\ref{sec:hadamard} describes how these samples are estimated on a quantum processor and defines the cost model.
Sec.~\ref{sec:cs} introduces the low-rank Hankel structure of the sampled signal and the compressed-sensing recovery used to achieve exponential speedup in evaluating QPD matrix elements.
Finally, Sec.~\ref{sec:protocol} assembles the complete protocol and locates each error source in the accuracy bound.

Throughout this work, we consider second-quantized electronic Hamiltonians $H$ with an active space of $n$ spatial orbitals
\begin{equation}
H=\sum_{pq}^{2n} h_{pq}\,a^\dagger_p a_q+\frac12\sum_{pqrs}^{2n} g_{pqrs}\,a^\dagger_p a^\dagger_r a_s a_q ,
\label{eq:Hel}
\end{equation}
where $a_p$ and $a
^{\dagger}_p$ are the annihilation and creation operators respectively acting on the spin-orbital $p$,
$h_{pq}$ are the usual one-electron integrals containing the kinetic energy and the electron-nuclei repulsion terms,
and $g_{pqrs}$ are the standard two-electron repulsion in chemist's notation
as described in Ref.~\cite{helgaker2013molecular}.
We denote the eigenstates of $H$, which solve the eigenvalue equation
$H\ket{\varphi_n}=E_n\ket{\varphi_n}$, by $\ket{\varphi_n}$,
and the corresponding eigenenergies by $E_0\le E_1\le\cdots E_m$ with $m \in \mathbb{N}_0$.
In addition, we define $\ket{\Phi}=U_\Phi\ket{\psi}$ as the initial state prepared by a particle-number-conserving unitary circuit $U_\Phi$ from a
computational-basis Slater determinant $\ket{\psi}$.
The overlaps $a_n$ of the Hamiltonian eigenstates with the initial state
are defined by $a_n=\braket{\varphi_n}{\Phi}$.

\subsection{Quantum Prolate Diagonalization}
\label{sec:pfd}

The original filter diagonalization\,\cite{wall1995extraction, mandelshtam1997low} method was developed from the insight that the quantum Hamiltonian eigenvalue problem can be recast as a signal recovery problem (see Ref.\,\cite{stroschein2026after} for a detailed discussion on the topic). In fact, it is easy to notice that the autocorrelation function $C(t)$ contains all the spectral information within a state $|\Phi\rangle$

\begin{equation}
C(t)=\bra{\Phi}e^{-iHt}\ket{\Phi}=\sum_n |a_n|^2 e^{-iE_n t},\qquad C(0)=1.
\label{eq:Csignal}
\end{equation}

Based on this correspondence, interested in studying only the low-lying portion of the Hamiltonian spectrum, we construct a subspace representation of $H$ by applying different spectral filter functions $\mathcal{F}_n(H)$ that faithfully reproduce the original spectrum of $H$ within the band of interest $B_f=[\omega^\ast-W_f,\omega^\ast+W_f]$, defined by a center $\omega^\ast$ and a half-width $W_f$.
Without loss of generality, we can center the band at zero by multiplying by a phase factor,
$C(t)e^{-i\omega^\ast t}$,
thus shifting the spectrum of the signal.

The Hamiltonian $H$ in these methods is described by the generalized eigenvalue problem (GEP) in the trial-vector space $\text{span}\{\mathcal{F}_n(H)|\Phi\rangle\}$.
Particularly, we write the filter functions $\{f_l\}_{l=0}^{M-1}$ (where $M$ is the dimension of the trial space) supported on $[-T,T]$ with Fourier transforms
$F_l(E)=\int_{-T}^{T} f_l(\tau)\,e^{-iE\tau}\,d\tau$.

Applying $F_l(H)$ to the initial state gives the filtered vectors
\begin{equation}
|v_l\rangle=F_l(H)\ket{\Phi}=\int_{-T}^{T} f_l(\tau)\,e^{-iH\tau}\ket{\Phi}\,d\tau
=\sum_k a_k F_l(E_k)\ket{\varphi_k}.
\label{eq:trial}
\end{equation}
Each eigenstate enters $|v_l\rangle$ with the weight $a_k F_l(E_k)$.
If $|F_l(E)|$ is small for $E\notin B_f$, the eigenstates outside the band are suppressed in every $|v_l\rangle$, and the $|v_l\rangle$ span, up to this leakage, the eigenstates inside $B_f$ that overlap with $\ket{\Phi}$.

Projecting the Hamiltonian onto each of these vectors we obtain the $M\times M$ generalized eigenvalue problem 
\begin{equation}
(V^\dagger HV)\,\mathbf b_i=\tilde E_i\,(V^\dagger V)\,\mathbf b_i,
\label{eq:GEP}
\end{equation}
where $\textbf{b}_i$ is the eigenvector of the subspace represented Hamiltonian in the filtered basis and $V$ is the matrix that collects all the vectors $|v_l\rangle$. 

The matrix elements read
\begin{align}
(V^\dagger HV)_{sl}&=-i\int_{-T}^{T}\!\!\int_{-T}^{T} f_s(\tau)\,\partial_\tau C(\tau-t)\,f_l(t)\,dt\,d\tau ,
\label{eq:VHV}\\
(V^\dagger V)_{sl}&=\int_{-T}^{T}\!\!\int_{-T}^{T} f_s(\tau)\,C(\tau-t)\,f_l(t)\,dt\,d\tau,
\label{eq:VV}
\end{align}
specified only in terms of the signal $C(t)$.

Eqs~\eqref{eq:VHV}--\eqref{eq:VV} are the matrix elements in their continuous-time form. For the purposes of our work it is important to recall the procedure developed in Ref.\,\cite{stroschein2026ground} to replace the double integrals by finite sums over samples $C(t_k)$ as this will be the starting point of our extension.

Before discussing the discretization procedure, it is worth clarifying that in QPD we choose a very specific set of filter functions that minimizes spurious leaking from out-of-band frequencies into the representation subspace. These functions are the time-limited prolate spheroidal wave functions\,\cite{slepian1961prolate, levitina2009filter, stroschein2026ground}. They can be defined as eigenfunctions $\{\xi_n^{(W)} \}_{n\geq0}$ of the finite-time and band-limited Fourier transform operator as in
\begin{equation}
\int_{-T}^{T} e^{\,i\frac{W\omega}{T}t}\,\xi^{(W)}_n(t)\,dt=\mu_n\sqrt{\frac{T}{W}}\,\xi^{(W)}_n(\omega),
\qquad \frac{|\mu_n|^2}{2\pi}=\gamma_n,
\label{eq:B5}
\end{equation}
with the time (frequency) domain spanning $[-T,T]$ ($[-W,W]$))
and decreasing sequence of coefficients $1>\gamma_0>\gamma_1>\cdots\to0$.
The eigenvalues depend on $W$ and $T$ only through $c=WT$, written $\gamma_n(c)$. The more the eigenfunctions $\xi^{(W)}_n$ are concentrated inside the window $W$, the closer the eigenvalues are to 1. 
As defined in Eq.\,\eqref{eq:B5} the prolates $\{\xi_n^{(W)} \}_{n\geq0}$ form a basis on $[-T,T]$ with $\int_{-T}^{T}\xi_n\xi_m\,dt=\gamma_n\delta_{nm}$. Finally, they also form a basis over the complete domain $(-\infty, \infty)$ for $W$-band-limited functions\,\cite{walter2003sampling}.

% ---------------------------------------------------------------------
\subsubsection*{Sampled PFD: matrix elements from equidistant samples}
\label{sec:sampled}

Even though Eqs.\,\ref{eq:VHV}--\ref{eq:VV} give an instruction to form the GEP, they are not very practical as they assume continuous access to the signal $C(t)$ which would make the computational cost of the algorithm extremely high. To overcome this issue, Ref.\,\cite{stroschein2026ground} provides a recipe for discretizing the double integrals in Eqs.\,\eqref{eq:VHV}--\eqref{eq:VV} in an error controlled manner. This result is achieved using the following interpolation formula:

\begin{equation}
f(t)\approx\frac{\pi}{W_s}\sum_{n=0}^{N_s}\;\sum_{k=-K}^{K} f\!\left(\tfrac{k\pi}{W_s}\right)\xi^{(W_s)}_n\!\left(\tfrac{k\pi}{W_s}\right)\xi^{(W_s)}_n(t),
\qquad K=\Bigl\lfloor\tfrac{W_sT}{\pi}\Bigr\rfloor,\; N_s=2K.
\label{eq:sampling}
\end{equation}
Where $N_s = 2K$ is the number of positive grid points $t_k$ with $k=1\dots N_s$ s.t. $t_k \in (0, T_{max}]$, with $T_{max} = 2T$.

For this interpolation formula to hold, the sampling rate $W_s$ must satisfy $W_s \geq W_C + W_f$, where $W_C$ is the bandwidth of the autocorrelation function $C(t)$. Convergence of this approximation requires the function $f(t)$ to be bandlimited with bandwidth at most $W_s$\,\cite{stroschein2026ground, walter2003sampling}. In our specific case, the functions $f(t)$ we wish to represent are the integrands in Eqs.\,\eqref{eq:VHV}--\eqref{eq:VV}. $f(t)$ is the product of a filter (or its derivative) and a shifted copy of $C(t)$. The former has a spectrum concentrated in $[-W_f, W_f]$, while the latter has a spectrum concentrated in $[-W_C, W_C]$. According to the convolution theorem, the spectrum of these integrands is therefore contained in $[-(W_C + W_f), W_C + W_f]$. The accuracy of this approximation is bounded by the concentration properties of the prolates, which are discussed in detail in Ref.\,\cite{stroschein2026ground}.

Applying Eq.~\eqref{eq:sampling} to an integral over $[-T,T]$ and using
Eq.~\eqref{eq:B5} at $\omega=0$,
$\int_{-T}^{T}\xi^{(W_s)}_n\,dt=\mu_n\sqrt{T/W_s}\,\xi^{(W_s)}_n(0)$,
yields an approximate discretization of the integral on the grid $t_k=\pi k/W_s$,
\begin{equation}
\int_{-T}^{T} f(t)\,dt\;\approx\;\sum_{k=-K}^{K} w_k\,f(t_k),
\qquad
w_k=\frac{\pi}{W_s}\sqrt{\frac{T}{W_s}}\sum_{n=0}^{N_s}\mu_n(c_s)\,\xi^{(W_s)}_n(0)\,\xi^{(W_s)}_n(t_k).
\label{eq:weights}
\end{equation}
The weights $w_k$ depend only on $(W_s,T)$ and are computed only once playing the role of the uniform trapezoidal weight $\delta t=\pi/W_s$ of a
naive Riemann discretization. It is important to note that, although we draw a parallel with integral discretization to provide further intuition, Eq.\,\eqref{eq:weights} is obtained through the analytical integration of Eq.\,\eqref{eq:sampling}. The approximation symbol arises from the approximation inherent in the interpolation formula.

Applying Eq.~\eqref{eq:weights} to both integrals of
Eq.~\ref{eq:VV}, and to Eq.~\ref{eq:VHV} after moving the derivative
onto the filter through integration by parts in $t$
(using $\partial_\tau C(\tau-t)=-\partial_t C(\tau-t)$), yields the
desired discretization. 

More specifically, let $\Toep\in\Cc^{(2K+1)\times(2K+1)}$ be the
Hermitian Toeplitz matrix of signal samples,
\begin{equation}
\Toep_{kk'}=C(t_k-t_{k'}),\qquad k,k'\in\{-K,\dots,K\},
\label{eq:toeplitz}
\end{equation}
which involves $C(t)$ at the $2N_s+1$ grid points $t_j$, $|j|\le N_s$,
i.e., on $[-2T,2T]$. Let $X,Y\in\Rr^{M\times(2K+1)}$ and $Z\in\Cc^{M\times(2K+1)}$ collect the filters
and their derivatives on the grid, $X_{lk}=\xi^{(W_f)}_l(t_k)$,
$Y_{lk}=\partial_t\xi^{(W_f)}_l(t_k)$, $Z_{lk}=C(t_k-T)\,\xi^{(W_f)}_l(T)
   -C(t_k+T)\,\xi^{(W_f)}_l(-T)$, and $\mathsf W=\mathrm{diag}(w_k)$.
Then
\begin{equation}
\begin{split}
B_{sl} = \sum_{k,k'} w_k w_{k'}\,
   \xi^{(W_f)}_s(t_k)\,
   C(t_k-t_{k'})\,
   \xi^{(W_f)}_l(t_{k'}) = \bigl(X\mathsf W\Toep\mathsf W X^{\!\top}\bigr)_{sl},
\end{split}
\label{eq:Bsampled}
\end{equation}

\begin{equation}
\begin{split}
A_{sl}
={}&-i\sum_{k,k'} w_k w_{k'}\,
   \xi^{(W_f)}_s(t_k)\,
   C(t_k-t_{k'})\,
   \partial_t\xi^{(W_f)}_l(t_{k'})
\\
&+i\sum_k w_k\,\xi^{(W_f)}_s(t_k)
 \Bigl[
   C(t_k-T)\,\xi^{(W_f)}_l(T)
   -C(t_k+T)\,\xi^{(W_f)}_l(-T)
 \Bigr]
\\
={}&-i\bigl(X\mathsf W\Toep\mathsf W Y^{\!\top}\bigr)_{sl}
   +i\bigl(X\mathsf W Z^{\!\top}\bigr)_{sl}.
\end{split}
\label{eq:Asampled}
\end{equation}
Eqs.~\eqref{eq:Bsampled}--\eqref{eq:Asampled} are the expressions of the GEP directly in terms of measured quantities: a vector of $2N_s+1$ complex numbers $C(t_j)$ enters only the Toeplitz matrix $\Toep$ and the boundary matrix $Z$, while everything else ($X,Y,\mathsf W$) is classical and signal-independent. We also note that the discretization of $VHV$, Eq.\,\eqref{eq:Asampled}, is Hermitian up to the discretization error introduced by Eq.\,\eqref{eq:sampling}. This could be explicitly enforced by replacing $A$ with $\tilde{A} = \frac{1}{2}(A + A^{\dagger})$. However, we did not find this necessary in our implementation, as the discretization error is always negligible compared to other sources of error such as shot noise, hardware noise and Trotter error.

We close this section by stating the accuracy bound derived in Theorem 2 of Ref.\,\cite{stroschein2026ground} in a form that is specialized for the ground state energy, which is the focus of this work:

\begin{equation}
    |\tilde E_0-E_0|\le\ \frac{\epsilon^{\text{Prlt}}_M}{\lambda_m(B_M^m) -   \epsilon^{\text{Prlt}}_M } \big[\sum_{E_k > W_f} (E_k - \tilde E_0) |a_k|^2  + \lambda^\ast_{\max}(\delta A^m_M-\tilde E_0\delta B^m_M)\big] \,\, .
    \label{eq:error_bound}
\end{equation}
$\lambda^*_{max}$ is the largest eigenvalue of the matrix sum in the parentheses, which is defined in terms of the error matrices $\delta A_M^m=A_M^m-V^\dagger HV$,
$\delta B_M^m=B_M^m-V^\dagger V$. Where $[A^m_M,B^m_M]$ is the GEP obtained by projecting the matrices of Eqs.\,\eqref{eq:Asampled}-\eqref{eq:Bsampled} onto the subspace formed by the $m$ leading singular values of $B_M$. 

With this the accuracy of the estimated eigenvalue depends on several distinct factors. (i) The intrinsic error of representing the Hamiltonian in the trial subspace is captured by $\epsilon^{\mathrm{Prlt}}_M$. This real and non-negative parameter depends on the trial space dimension $M$ and on $c=W_f T$. For the values of $M = \frac{c}{\pi}$ adopted in this work it satisfies $0 \leq \epsilon^{\mathrm{Prlt}}_M\leq 2\pi M T(1+c)[1 - \gamma_{M-1}(c)]$ where $1 - \gamma_{M-1}(c)$ is related to the norm of $\xi^{W_f}_{M-1}$ out of the band $B_f$. Its value decays exponentially with $T_{max}$ (cfr. Eq. 31, Ref.\,\cite{stroschein2026ground}). (ii) The quality of the initial state preparation affects the estimation of the in-band frequency with a factor that grows linearly with the squared overlap $|a_k|^2$ of the out-of-band eigenstates. (iii) Any further error affecting the sampled matrices is captured inside $\delta A^m_M$ and $\delta B^m_M$. These matrices collect \emph{every} deviation between the actually constructed sampled matrices actually and the exact continuous ones. In Ref.~\cite{stroschein2026ground} only the prolate-sampling truncation error and the shot noise has been included. In the present work $\delta A^m_M,\delta B^m_M$ additionally contain the quantum hardware noise, the reconstruction error of Sec.~\ref{sec:cs}, and the Trotter error. We will come back to this discussion on the different factors affecting the accuracy in Sec.\,\ref{sec:results}.

% ---------------------------------------------------------------------
\subsection{Measuring autocorrelation function samples with a Hadamard test}
\label{sec:hadamard}

Samples of the real ($\Re$) and imaginary ($\Im$) parts
of the autocorrelation function $C(t_k)$
can be obtained with a quantum computer by the Hadamard test circuit\,\cite{cleve1998quantum}.

\begin{equation}\label{eq:htest_circ}
    \Re/ \Im [ \langle \Psi | U_H^{(t_k)} | \Psi \rangle ]
    =
    \Qcircuit @C=1em @R=1em {
        & \qw & \gate{H} & \ctrl{1} & \gate{I/S} & \gate{H} & \meter \\
        \lstick{|x_0\rangle} & {/}^{\otimes^n} \qw & \gate{U_\Phi} & \gate{\rule[0em]{0pt}{1em} U_H^{(t_k)}} & \qw & \qw & \qw
    }
\end{equation}

Initially, a Hadamard gate is applied to the ancilla qubit,
and the initial state $\ket{\Phi}=U_\Phi\ket{x_0}$ is prepared on the state
register. Next, the time-evolution circuit $U^{(t_k)}_H = e^{-iHt_k}$,
controlled on the ancilla, is applied to the system register.
After applying a final Hadamard gate on the ancilla, the Hadamard test circuit
prepares the quantum state 

\begin{equation*}
\tfrac12\bigl[\ket0\otimes(\ket\Phi+U^{(t_k)}_H\ket\Phi)+\ket1\otimes(\ket\Phi-U^{(t_k)}_H\ket\Phi)\bigr]\ ,
\end{equation*}
and the measurement outcomes $c_x\in\{+1,-1\}$ on the ancilla
qubit are

\begin{equation}\label{eq:Hadamard-re}
P[c_x(t_k)=\pm1]=\tfrac12\bigl(1\pm \Re ( \,C(t_k)) \bigr).
\end{equation}
By inserting a phase gate $S=\mathrm{diag}(1,i)$ on the ancilla before application
of the final Hadamard gate, one instead obtains measurement
outcomes $c_y\in\{+1,-1\}$ of

\begin{equation}\label{eq:Hadamard-im}
 P[c_y=\pm1]=\tfrac12(1\pm\mathrm{Im}\,C(t_k)) \ ,   
\end{equation}
which are related to the imaginary part of the autocorrelation function.

From Eqs. \eqref{eq:Hadamard-re} and \eqref{eq:Hadamard-im},
the autocorrelation function is reconstructed as
$C(t_k) = \mathbb E[c_x+ic_y]$.
By averaging over $\Nshots$ outcomes of each circuit, the unbiased estimator
\begin{equation}
\tilde C(t_k)=C(t_k)+\nu(t_k),
\label{eq:estimator}
\end{equation}
is obtained, where $\nu(t_k)$ denotes the statistical (shot) noise.
As shown in Appendix A of Ref.\,\cite{ding2023even}, to keep $|\nu(t_k)|\le\eps$ for all $N_s$ positive grid points with probability $\ge1-\delta$, a number
of shots per circuit of $\Nshots=O(\eps^{-2}\log(N_s/\delta))$ is required.

Importantly, the time evolution operator
employed in our noisy-quantum-hardware and emulator
experiments is not the exact operator $U_H^{(t_k)}$,
but rather an approximation obtained by Trotterization, as detailed in Appendix\,\ref{sec:trotter}.
The associated Trotterization error is coherent,
grows with $t_k$, and is not considered in Eq.~\eqref{eq:estimator}.
Throughout the numerical experiments that follow, we will state explicitly on a case-by-case basis, whether or not the reference against which energies are compared is the result from exact diagonalization or noiseless execution of the Trotterized circuit. Only the latter isolates the effect of hardware noise from the product-formula bias.

Two quantities that are central to the comparisons in our numerical studies
are (i) the total evolution time executed on the device $T_{runtime}$,
and (ii) the largest evolution time $\Tmax$, for which we sample
the autocorrelation function. We define the total runtime as
\begin{equation}
T_{runtime}=\Nshots\sum_{k\in\mathcal K} t_k ,
\label{eq:Trun}
\end{equation}
where $\mathcal K$ is the set of considered positive-time indices.
For standard QPD, $\mathcal K=\{1,\dots,N_s\}$. For the compressed-sensing variant of Sec.~\ref{sec:cs} (see below), $\mathcal K=\Om$ with $|\Om|=O(\log N_s)$.

The largest evolution time, following the notation of the interpolation formula in Eq.\,\eqref{eq:sampling}, is defined as
\begin{equation}
 \Tmax=\pi N_s/W_s \ ,
 \label{tmax}
\end{equation}
which is a proxy for circuit depth, required
coherence time, and the number of distinct circuits $|\mathcal K|$.
Note that Eq.~\eqref{eq:Trun} only tracks the total Hamiltonian
simulation time; it does not account for gate counts per unit time,
which depend on the compilation of $U_H^{(t)}$.
Where relevant, we will report on these separately.

We set the number of shots per circuit (for both QPD and sparse QPD) to be
\begin{equation}
\Nshots=F\sqrt{N_s\log N_s} \ ,   
\end{equation}
with a constant $F$. According to standard results in random matrix theory (e.g. Ref.\,\cite{vershynin2012compressed}), the norm of the perturbation matrix $||\delta\Toep||_{sn}$ due to shot noise is $||\delta\Toep||_{sn} = \mathcal{O}(\sqrt{\frac{M}{N_{shots}}})$. Substituting our choice of $M=W_f T_{max}/\pi$, and using Eq.\,\ref{tmax}, we have up to logarithmic factors $||\delta\Toep||_{sn} = \mathcal{O}(T_{max}^{\frac{1}{4}})$.

\subsection{Compressed sensing of the sampled signal}
\label{sec:cs}

Sec.\,\ref{sec:pfd} showed that quantum computer measurements enter the GEP of Eq.\,\eqref{eq:Bsampled}-\eqref{eq:Asampled} through $N_s$ samples of $C(t)$. Here, we show how the number of measured samples can be reduced from $N_s$ to $\mathcal{O}(\text{log}(N_s))$ while the grid, and with it the matrices of Eq.\,\eqref{eq:Bsampled}-\eqref{eq:Asampled}, are left unchanged. To this end, we measure $C(t)$ on a random subset of the grid and reconstruct the remaining samples classically. The reconstruction exploits the fact that, when the initial state has most of its weight on a few eigenstates of $H$, the samples can be arranged into a Hankel matrix close in Frobenius norm to a low rank matrix. 

This reformulation of signal recovery as matrix completion is an established approach~\cite{candes2010matrix,candes2010power}; here, we use the recently developed Hankel-structured Newton-like descent (HSNLD) algorithm of Ref.~\cite{Cai2025}. We focus on the main ingredients required to derive the new algorithm and its error analysis, and defer the algorithmic details and the recovery guarantee to  Appendix~\ref{app:hsnld}.

\subsubsection*{Low-rank structure of the sampled signal}

We gather the grid samples into the vector $\mathbf x=(x_1,\dots,x_N)^{\top}\in\Cc^N$, with $N=2N_s+1$ and $x_j=C(t_{j-1-N_s})$, so that $\mathbf x$ covers the interval $[-\Tmax,\Tmax]$ with spacing $\pi/W_s$. Inserting Eq.~\eqref{eq:Csignal} gives
\begin{equation}
x_j=\sum_n |a_n|^2\,z_n^{\,j-1-N_s},\qquad z_n=e^{\,-i\pi E_n/W_s},
\label{eq:sum-exp}
\end{equation}
 
The Hankel matrix is derived from $\mathbf x$ as
\begin{equation}
\Hank(\mathbf x)=
\begin{pmatrix}
x_1 & x_2 & \cdots & x_{N_2}\\
x_2 & x_3 & \cdots & x_{N_2+1}\\
\vdots & & & \vdots\\
x_{N_1} & x_{N_1+1} & \cdots & x_N
\end{pmatrix}\in\Cc^{N_1\times N_2},\qquad N_1+N_2-1=N ,
\label{eq:hankel}
\end{equation}
whose columns are sliding windows of length $N_1$ over $\mathbf x$. Each element is defined as $\Hank(\mathbf{x})_{ik} = \textbf{x}_{i+k-1}$. If
Eq.~\eqref{eq:sum-exp} contains $r$ distinct frequencies, then
$\Hank(\mathbf x)$ admits the Vandermonde
decomposition~\cite{boley11998vandermonde, liao2016music}
$\Hank(\mathbf x)=E_L\,D\,E_R^{\top}$ with $(E_L)_{i,n}=(E_R)_{i,n}=z_n^{\,i-1}$ and
$D=\mathrm{diag}(|a_n|^2z_n^{-N_s})$, so that $\mathrm{rank}\,\Hank(\mathbf x)=r$ as soon as $N_1,N_2\ge r$. 

For a molecular Hamiltonian, the number of eigenstates with nonzero overlap is on the order of the Hilbert-space dimension, so $\Hank(\mathbf x)$ is not exactly low-rank. In fact, considering noisy samples, the matrix will be in general full rank. What makes reconstruction possible is that, for most molecules relevant to chemistry, it is possible to prepare initial states for which $|a_0|^2\gg\sum_{n>0}|a_n|^2$\cite{erakovic2025high, fomichev2024initial}. In this case, the vast majority of the amplitudes are also very small, with magnitudes on the order of the inverse of the Hilbert-space dimension, making them negligible even compared with the shot noise. The completion rank $r$ required by the HSNLD algorithm must account for every component whose weight exceeds the noise level of the data, it corresponds to the number of exponentials used to best fit the data of Eq.\,\eqref{eq:sum-exp}. It depends on the overlaps $a_n^2$ and on the frequency gap that we can actually resolve. A sufficient condition for $r=1$ (as employed throughout this work) is that the residual weight $1 - a_0^2$ lies below the noise level. When $r \neq r_{\mathrm{true}}$, an additional model error is introduced. Since estimating the correct rank is generally challenging, we can instead choose $r \geq r_{\mathrm{true}}$ at the cost of increasing the number of samples, relying on the prolate filtering step to suppress noise-associated components.

\subsubsection*{Sampling strategy}

One of the goals of this study is to show that, assuming efficient state preparation, for the accuracy needed in quantum chemistry applications, it is possible to decouple the accuracy of the energy estimates from the length of the sampled signal up to logarithmic factors. 

Regarding QPD, the window $[-T_{max}, T_{max}]$ sets the number of frequencies that the GEP can resolve as $n_{freq} \leq \frac{W_{f} T}{\pi}$, growing linearly in $T_{max}$\,\cite{stroschein2026ground}. Likewise, the recovery guarantees of off-the-grid compressed sensing assume a frequency separation of order $T_{max}^{-1}$ for randomly selected samples\cite{tang2013compressed} and $\log(T_{max})^{-1}$ for consecutive samples~\cite{candes2014towards}. 

However, if the minimum frequency gap within $B_f$ can be controlled by the state preparation routine, one can reduce the required $T_{\max}$ for the completion protocol to be successful. By allocating more resources to the quantum state preparation routine and a classical optimization problem, both the circuit depth and runtime can be reduced. Since quantum state preparation is generally expected to be less costly than long time evolutions~\cite{erakovic2025high, fomichev2024initial,chen2023efficient}, we expect this allocation to be beneficial. Moreover, given that the dimension of the classical optimization problem is much smaller than that of the Hilbert space typically handled in quantum simulations, the classical computational overhead will be negligible compared to the quantum cost. Clearly, we do not expect this assumption to hold in general. For strongly correlated systems with significant multireference character or systems close to the metallic regime, the trade-off between more costly state preparation and longer time evolutions deserves further investigation. However, in such a case, the standard classical reference (i.e. CCSD(T)\,\cite{raghavachari1989fifth}) would also struggle.

Let $\Om\subset\{1,\dots,N_s\}$ be the set of positive-time indices executed on the quantum processor. Together with their mirror images, obtained from $\tilde C(-t_k)=\tilde C(t_k)^\ast$, and with the known value $C(0)=1$, they define the set $\Om^{\pm}\subset\{1,\dots,N\}$ of observed entries of $\mathbf x$, with $|\Om^{\pm}|=2|\Om|+1$. Let $\PiOm$ be the operator that keeps the entries of a vector in $\Om^{\pm}$ and sets the others to zero. The data available to the reconstruction are
\begin{equation}
\textbf{f}=\PiOm\tilde{\mathbf x},\qquad \tilde{\mathbf x}=\mathbf x+\boldsymbol\nu ,
\label{eq:obs}
\end{equation}
where $\tilde x_j=\tilde C(t_{j-1-N_s})$ is the estimator of Eq.~\eqref{eq:estimator}, of which only the entries in $\Om^{\pm}$ are measured, and $\boldsymbol\nu$ collects its deviation from the exact signal. In noiseless simulations $\boldsymbol\nu$ is the shot noise; on hardware it also contains the effects of hardware noise and Trotter error (Sec.~\ref{sec:protocol}).

The recovery algorithm of Ref.\,\cite{Cai2025} requires for a signal of rank $r$ a number of sampled points $|\Omega| \propto r^2 \text{ln}(N_s)$. In this work, we set the number of sampled points as
\begin{equation}
|\Om|=C_{\mathrm{cs}}\ln N_s,
\label{eq:keep}
\end{equation}
with a constant $C_{\mathrm{cs}}$ absorbing all the other algorithm-specific parameters discussed in Appendix \ref{app:hsnld}.

The indices in $\Om$ are drawn without
replacement according to a density $q_k$ over $k=1,\dots,N_s$. We consider three densities, which differ in how strongly they favor short evolution times:

\begin{equation}
q_k\propto
\begin{cases}
1 ,\\
\exp(-\beta\,t_k/\Tmax),\\
\exp\!\bigl(-t_k^2/2\sigma^2\bigr).
\end{cases}
\label{eq:densities}
\end{equation}
The exponential density has the decay parameter $\beta$, the half-normal density is inspired by Ref.~\cite{ding2024quantum} and its width $\sigma$ is fixed by requiring that the expected largest executed time be $\mathbb E[\max_{k\in\Om}t_k]=\tau\Tmax$, with $\tau\in(0,1]$. For the two biased densities, the largest executed time

\begin{equation}
T_{max}^{\Omega}=\max_{k\in\Om}t_k
\label{eq:tmax-omega}
\end{equation}
is in general smaller than $T_{max}$, and it is $T_{max}^{\Omega}$ that sets the depth of the deepest circuit.
While a bias towards short times further reduces $T_{max}$ as well as $T_{runtime}$, and under hardware noise favors samples less attenuated by decoherence, at the same time it shrinks the effective observation window and the consequent frequency resolution. We will assess the impact of these different effects numerically in Sec.\,\ref{sec:results}.

\subsubsection*{Hankel matrix reconstruction}

Given $\mathbf f$, the reconstruction task is to find the vector
$\hat{\mathbf x}$ whose Hankel matrix has rank $r$ and which fits the observed entries. Since an unobserved $x_j$ removes an
entire skew-diagonal of $\Hank(\mathbf x)$, the redundancy of the Hankel
structure alone cannot restore it and the rank constraint must be enforced
explicitly~\cite{Cai2025}. HSNLD does so by writing the Hankel matrix as a product $LR^\ast$ of two rank-$r$ factors, $L\in\Cc^{N_1\times r}$,
$R\in\Cc^{N_2\times r}$, and minimizing
\begin{equation}
\ell(L,R)=
\underbrace{\frac{1}{2p_{\Om}}\sum_{j\in\Om^{\pm}}\varsigma_j\,\bigl|[\Hank^{\dagger}(LR^\ast)]_j-\tilde x_j\bigr|^2}_{\text{deviation from observed entries}}
+\underbrace{\frac12\bigl\|LR^\ast-\Hank\Hank^{\dagger}(LR^\ast)\bigr\|_F^2}_{\text{deviation from Hankel structure}} .
\label{eq:objective-main}
\end{equation}
Here, $p_{\Om}=|\Om^{\pm}|/N$ denotes the fraction of observed entries, and $\Hank^{\dagger}$ maps an $N_1\times N_2$ matrix to its skew-diagonal averages (it is the pseudo-inverse of the map $\Hank$ defined in Eq.~\eqref{eq:hankel}). The quantity $\varsigma_j=\min\{j,N_1,N_2,N-j+1\}$ gives the number of entries on the $j$-th skew-diagonal. With these weights, the first term equals the squared Frobenius norm of the difference between the model and data Hankel matrices, restricted to the observed skew-diagonals. Since $LR^\ast$ is not necessarily Hankel, this structure is enforced through the regularization term. In fact, $\Hank\Hank^{\dagger}$ is the orthogonal projection onto the space of Hankel matrices.
 
HSNLD minimizes Eq.~\eqref{eq:objective-main} by gradient steps on $L$ and $R$, in which the two gradient components are multiplied by the $r\times r$ matrices $(R^\ast R)^{-1}$ and $(L^\ast L)^{-1}$, respectively (see Eq.~\eqref{eq:hsnld-step} for the gradient update rule in the SI). This preconditioning makes the convergence rate independent of the condition number of $\Hank(\mathbf x)$\,\cite{tong2021accelerating} . Every operation acts on the $N$
distinct entries of the Hankel matrix rather than on the full array, so
one iteration costs $\mathcal O(rN\log N+r^2N)$ operations, and the
iteration is started either from the best rank-$r$ approximation of the
zero-filled data or from a previous estimate of the
signal when one is available. The output is the completed vector
$\hat{\mathbf x}$, i.e. an estimate $\hat C(t_j)$ for every grid point
$|j|\le N_s$.

It is important to note that the algorithm incorporated into standard QPD in this work is guaranteed to converge linearly only under assumptions that are not necessarily satisfied in our setting. In particular, all existing proofs of convergence assume uniform random sampling, whereas the other sampling distributions considered in this work have not been formally studied. Moreover, all the proofs in Ref.\,\cite{Cai2025} assume noiseless access to the sampled signal, with the exception of a fraction \(\alpha\) of outliers. In our setting, however, the dataset is not necessarily corrupted by outliers; rather, each sample is affected by noise $\nu(t_k)$, which may arise from a finite number of measurement shots, the use of a quantum simulation algorithm, or hardware-related parameters. From our numerical experiments, however, we observed that these deviations were not critical.

% ---------------------------------------------------------------------
\subsection{Sparse QPD algorithm}
\label{sec:protocol}
Finally, we present the steps of sparse QPD in Algorithm~\ref{alg:csq-pd}. Once the sampling rate has been estimated (e.g., either from norm bounds or through inexpensive classical calculations, see Sec.\,\ref{sec:pfd}), the grid spacing is fixed. By choosing the number of positive grid points $N_s$, we also determine $T_{max}$, provided that the condition $n_{freq} \leq W_f T_{max}/\pi$ is satisfied. We note that the choice of a filter center $\omega^*$ (e.g., based on a Hartree--Fock calculation) is required.

These quantities, together with the sampling density, constitute the user-specified inputs to sparse QPD. All other parameters are determined according to the procedures discussed in the previous sections.

\begin{algorithm}[H]
\caption{Sparse QPD}
\label{alg:csq-pd}
\begin{algorithmic}[1]
\State Fix $W_s\ge W_C+W_f$, $N_s$; set $t_k=\pi k/W_s$, . Fix $\omega^*, W_f$ from classical calculations.
\State Draw $\Omega\subset\{1,\dots,N_s\}$ with $|\Omega|$ from Eq.~\eqref{eq:keep} and density $q_k$ from Eq.~\eqref{eq:densities}.
\State For each $k\in\Omega$ run the two Hadamard-test circuits (real, imaginary) with $N_{\mathrm{shots}}$ shots; form $\tilde C(t_k)$, Eq.~\eqref{eq:estimator}. \Comment{quantum step; cost $T_{\mathrm{run}}=N_{\mathrm{shots}}\sum_{k\in\Omega}t_k$}
\State Assemble $\mathbf{f}$, Eq.~\eqref{eq:obs}, using $\tilde C(-t_k)=\tilde C(t_k)^\ast$ and $C(0)=1$.
\State Run HSNLD, Eqs.~\eqref{eq:init}--\eqref{eq:hsnld-step}, at rank $r$ to obtain $\hat{\mathbf{x}}=\hat C(t)$.
\State Shift, $\hat C(t_j)\to\hat C(t_j)e^{-i\omega^\ast t_j}$; form $\Toep$, Eq.~\eqref{eq:toeplitz}, and $A,B$, Eqs.~\eqref{eq:Bsampled}--\eqref{eq:Asampled}, at trial dimension $M=\lceil \frac{W_f T_{max}}{\pi} \rceil$.
\State Detect $m$ from the spectrum of $B$, refine to $[A^m_M,B^m_M]$, solve the GEP; return $\tilde E_0+\omega^\ast$.
\end{algorithmic}
\end{algorithm}

Standard QPD is Algorithm~\eqref{alg:csq-pd} with $\Om=\{1,\dots,N_s\}$
and step\,5 omitted. The two protocols share the classical
post-processing (steps 6--7) exactly and their quantum computational cost differ by the ratio $\sum_{k\in\Om}t_k/\sum_{k=1}^{N_s}t_k$.

We conclude our theoretical discussion by examining how matrix completion errors affect the QPD error bound in Eq.\,\eqref{eq:error_bound}. Defining $\delta \textbf{x} = \hat{\textbf{x}} - \textbf{x}$ as the deviation between the completed and exact signals, $C(t)$, the corresponding deviation in $B$ follows from Eq.\,\eqref{eq:Bsampled} as
$\delta B_{m}^{M} = P^{M\dagger}_m X W \delta \Toep W X^T P^{M}_m$,
where $P_m^M$ projects the matrix onto the leading $m$-dimensional subspace. An analogous expression for $\delta A$ follows from Eq.\,\eqref{eq:Asampled}.

In the following sections, we investigate the performance of sparse QPD in the presence of errors due to shot noise, model error, and optimization error, none of which increase in magnitude with increasing $t$. Subsequently, we also consider hardware noise and Trotter error, both of which grow as $t$ increases.

\section{Results}
\label{sec:results}

\subsection{Computational details}
\label{sec:methods}

\begin{figure}[h!]
    \centering
    \includegraphics[width=0.85\linewidth]{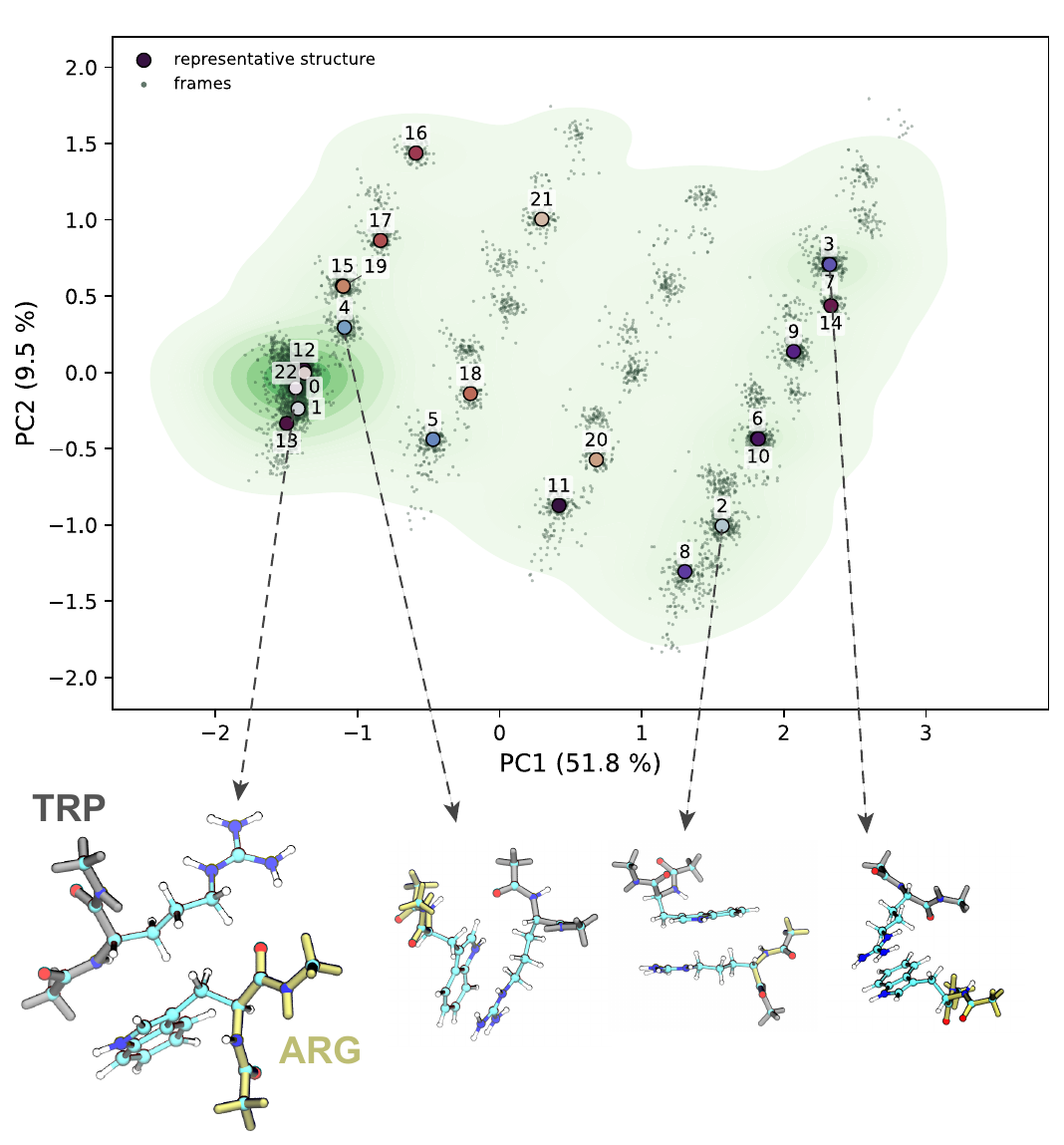}
    \caption{Representative structures of the arginine-tryptophan pair.
Top: umbrella sampling frames projected onto the first two principal components of a principal component analysis performed on the discretized feature space used for clustering. Individual frames are shown as dots and their density in light green; the representative structure of each cluster is labeled by its index, with clusters ranked by population. Bottom: four cluster representatives illustrating the structural diversity of the pair, with the arginine backbone in yellow and tryptophan backbone in grey.}
    \label{fig:clustering}
\end{figure}

In this section we evaluate Algorithm \ref{alg:csq-pd} on 23 structures of an arginine-tryptophan pair,
selected by clustering~\cite{descamps2025clustering} as the most representative configurations of a trajectory produced using molecular-dynamics umbrella sampling (see Fig.\,\ref{fig:clustering}). 
Additional details on the system and on the production of its representative interacting structures are provided in Appendix \ref{app:molecular}. Our goal is to assess sparse QPD in a weak correlation case
in which quantum computation may still speed up high-throughput
computational chemistry workflows.
By accounting for shot noise, we demonstrate that sparse sampling significantly reduces computational resources compared to standard QPD,
without significantly increasing circuit depth (see Fig.\,\ref{fig:density_comparison}).
We choose active spaces of up to three spatial orbitals and four electrons (3o, 4e),
to compare directly with the quantum hardware results (see below),
and in Fig.\,\ref{fig:scaling_active_space},
we analyze how the cost of the algorithm scales with the size of the active space up to (8o, 8e).
Both QPD and sparse QPD were run on a classical emulator and on the Quantinuum Helios quantum computer (Figs.\,\ref{fig:signal_recovery}-\ref{fig:energy_distribution}).
The algorithm scaling in this context is analyzed in Figs.\,\ref{fig:noise_scaling_small_active_space}-\ref{fig:noise_scaling_large_active_space},
with two error sources beyond shot noise,
namely hardware noise and Trotter error (see Sec.\,\ref{sec:trotter}).

We consider both bare, i.e. solvent free, complete active space configuration interaction (CASCI) Hamiltonians as described by Eq. \eqref{eq:Hel}, as well as a point charge embedding
QM/MM CASCI Hamiltonian $H_{\mathrm{QM/MM}}$, where the point charges of surrounding water molecules in the snapshot of the MD simulation affect the one-electron integrals:
\begin{equation}\label{eq:H_qmmm}
H_{\mathrm{QM/MM}} = H + \sum_{pq} h^{(\mathrm{QM/MM})}_{pq} a^\dagger_{p} a_q \ ,
\end{equation}
with the contribution $h_{pq}^{(\mathrm{QM/MM})}$ to the one-electron integrals being
\begin{equation}\label{h_qmmm_first_quant}
h_{pq}^{(\mathrm{QM/MM})} = \langle p| \sum_A^{N_{\mathrm{MM}}}\sum_B^{N_{\mathrm{QM}}}
\frac{Q_A Q_B}{\lvert R_A - R_B \rvert} +
\sum_{A}^{N_{\mathrm{MM}}}\sum_{i}^{N_{\mathrm{elec}}} \frac{Q_A}{\lvert R_A - r_i \rvert}|q\rangle \ ,
\end{equation}
where $p,q$ are the one-electron spin-orbitals, $N_{\mathrm{MM}}$ are the point charges in the MM region,
$N_{\mathrm{QM}}$ are the point charges in the QM region, and $N_{\mathrm{elec}}$ are the electrons in the QM region. $R_A$ labels the Cartesian coordinates of point charge $A$, $r_i$ the Cartesian coordinates of electron $i$.

While the embedded Hamiltonian was utilized for the quantum hardware calculation
and emulation performed in section \ref{quantinuum_results},
with details on the employed molecular structure and the point charge
model described in Appendix \ref{sec:qm_mm_sys},
all other (including shot noise only) QPD calculations discussed throughout
section \ref{sec:sparse_sampling_qpd} employed the bare CASCI Hamiltonians.
The full molecular structure production procedure can be found in Appendix~\ref{sec:struc_extraction}.
To analyze the algorithm's scaling properties, solvent effects are negligible provided they do not alter the ground-state overlap regime of interest, i.e., $a_0^2 \gg \sum_n a_n^2$. This condition holds for our structures, allowing us to perform the scaling analysis using water-free Hamiltonians. Nevertheless, we performed the quantum hardware experiments using the full Hamiltonian, including solvation effects, to ensure a realistic experimental validation of our implementation.

The orbital active spaces are constructed for both the bare and the embedded CASCI Hamiltonians as follows: the PySCF program package~\cite{sun2018pyscf} performs a Hartree-Fock (HF) and a subsequent MP2 calculation to obtain MP2 natural orbitals. Orbital active spaces are constructed around the MP2 HOMO and LUMO orbitals. For example, an active space of 3 orbitals and 2 electrons contains the MP2 HOMO and LUMO orbitals, as well as the LUMO+1 virtual orbital. All the calculations use a cc-pVDZ basis set~\cite{dunning1989gaussian}.

To enable straightforward Trotterized Hamiltonian simulation on quantum hardware, the Hamiltonians in Eq.~\eqref{eq:Hel} and~\eqref{eq:H_qmmm} are mapped to qubit Hamiltonians by a Jordan-Wigner transformation~\cite{jordan1928uber}.
For the quantum hardware calculations and emulations in section~\ref{quantinuum_results},
qubit tapering techniques~\cite{bravyi2017tapering} were
employed additionally to reduce the complexity of the qubit Hamiltonian.
Details on the Trotterization scheme and quantum circuit compilation are provided in Appendix~\ref{sec:trotter_scheme_det}.

In all the calculations below (i.e., for both QPD and sparse QPD), we set the initial state to $|\Phi\rangle = |HF\rangle$, corresponding to the Hartree-Fock determinant. The sampling rates were set to $W_S = 5$ and $W_f = 0.85$. We set the trial dimension $M$ equal to half of the essential dimension, $M=\frac{W_f T}{\pi}$, as indicated in Ref.\,\cite{stroschein2026ground}. Given the high overlap between the initial state and the ground state (see Appendix\,\ref{sec:overlaps}), we set the number of frequencies to be detected within $B_f$ to $m=1$ throughout. For the calculations performed on real quantum hardware (Sec.\,\ref{sec:results}), we set $m=2$ because, upon inspection of the spectrum of the matrix $B$ (see Eq.\,\eqref{eq:Bsampled}), we identified two dominant modes. As mentioned in Sec.\,\ref{sec:hadamard}, the number of shots is $\Nshots=F\sqrt{N_s\log N_s}$ with $F=1$. For the calculation on the Helios machine, we set $\Nshots = 10$.  

All the calculations presented in our scaling analysis (i.e., Figs.\,\ref{fig:density_comparison}, \ref{fig:scaling_active_space}, \ref{fig:noise_scaling_small_active_space}, and \ref{fig:noise_scaling_large_active_space} below) were repeated 10 times for each structure and value of $T_{\max}$. In the plots, the faint lines (or dots) represent the averages over the 10 repetitions, while the brighter lines are obtained by computing the median across the set of averages corresponding to the different structures.

Regarding settings specific to sparse QPD, we recall that $T_{\max}$ is defined here as $T_{\max} = \max{t_k} \in \Omega$. The compression factors used for the sparse QPD runs differ among the different figures. When studying the effect of the different sampling distributions in Fig.\,\ref{fig:density_comparison}, we set $\beta = 3$ for the exponentially biased sampling. For the truncated Gaussian sampling, we chose different values of $\sigma$ depending on the specific study; these values will be specified below when needed. The number of sampled points was set according to Eq.\,\eqref{eq:keep}. For Gaussian sampling, we set $C_{cs}=3$, while for the other sampling schemes, $C_{cs}=2$. For the numerical experiments presented in this paper, we did not observe a strong dependence of the results on the prefactor $C_{cs}$. We will discuss this point more in detail in the following in light of the formal details of the recovery algorithm discussed in Sec.\,\ref{app:hsnld}. 
For all calculations, we used the compressed-sensing recovery settings described below. The termination conditions for the HSNLD optimization were either a maximum number of iterations of 800 or a tolerance threshold of $10^{-8}$, with a learning rate of $\eta=0.5$. As mentioned above we did not consider the presence of sparse outliers in the signal since shot and hardware noise spectra densely cover the frequency domain. All the other parameters were chosen as indicated in Ref.\,\cite{Cai2025}. The scaling analysis in the presence of hardware noise was carried out using a noise model implemented through the software package PECOS\,\cite{pecos}. The parameters for the Quantinuum Helios machine were fetched from Ref.\,\cite{heliosparams}. We validated the reliability of our simulations using the offline noise model by comparing the results of randomized benchmarking experiments (see Appendix\,\ref{app:helios}) with those obtained using the Quantinuum in-house hardware emulator and the actual quantum hardware.

\subsection{Sparse sampling accelerates quantum prolate diagonalization}\label{sec:sparse_sampling_qpd}

\begin{figure}[h!]
        \centering
        \includegraphics[scale=0.85]{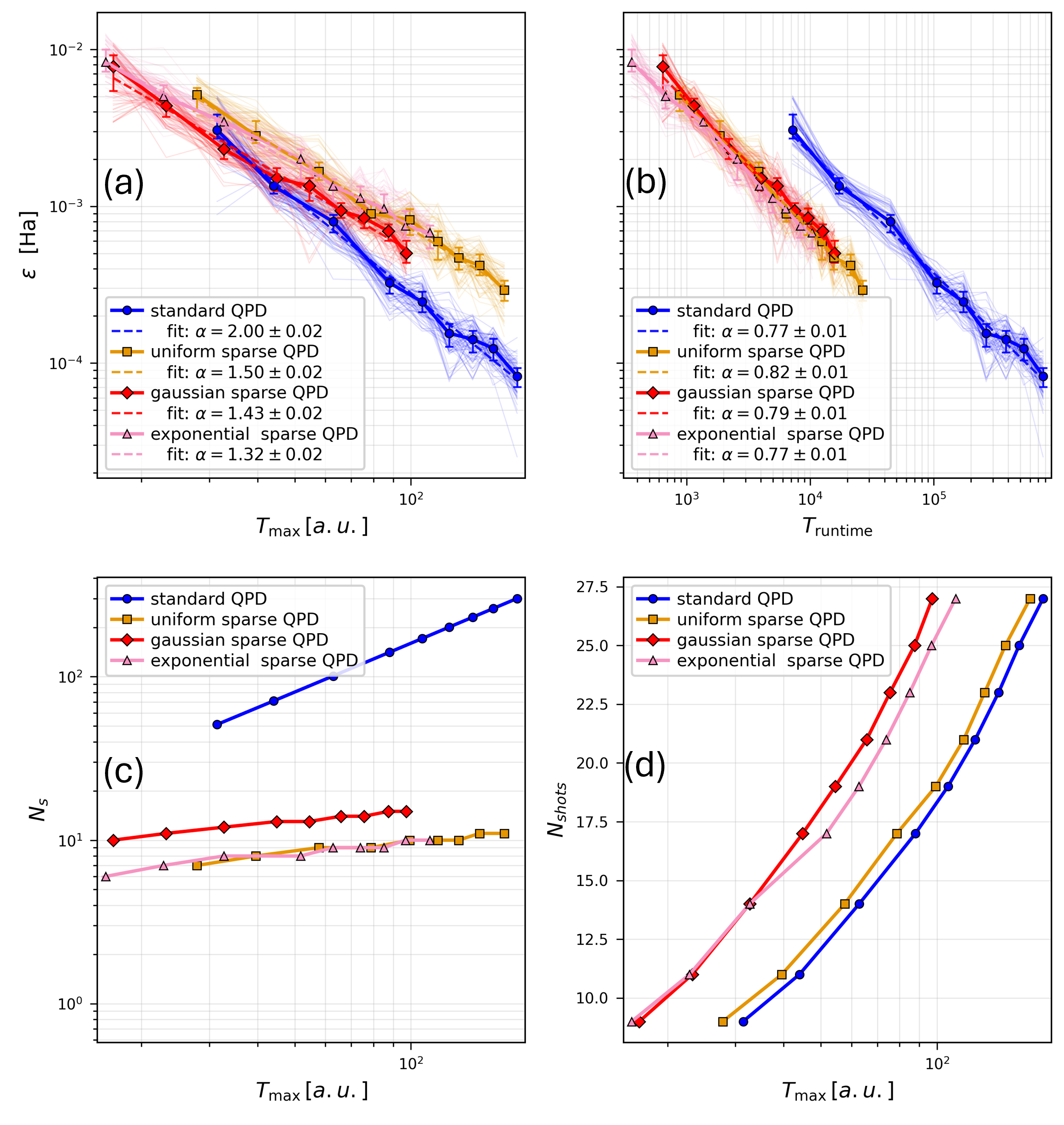}
        \caption{Sparse sampling reduces the computational cost of QPD. Ground state energy error as a function of the maximal evolution time $T_{max}$ (a) and the total runtime $T_{runtime}$ (b), as defined in Eq.~\eqref{eq:Trun}. Faint solid lines are obtained by averaging over 10 realizations of a (sparse) QPD calculation for a given value of $T_{max}$ and correspond to the different ARG-TRP geometries, as explained in the computational details section\,\ref{sec:methods}. Thick solid lines show the median values computed from the estimates for the different geometries, while the error bars correspond to the interquartile ranges. Number of sampled points $N_s$ (c), and number of shots per sample $N_{shots}$ (d) as a function of the maximal evolution time for the different methods and sampling densities studied.}
        \label{fig:density_comparison}
\end{figure}

First, we illustrate the effects of signal recovery on the QPD algorithm. In Fig.\,\ref{fig:density_comparison}, we show the ground state error of QPD (blue dots, solid blue line) compared to sparse QPD adopting different sampling distributions (Eq.\,\eqref{eq:densities}). As is apparent in Fig.\,\ref{fig:density_comparison}b sparse sampling allows us to reduce the $T_{runtime}$ by approximately one order of magnitude to achieve the same level of accuracy. Such a speed up is reflected in the number of sampled points needed to achieve an error estimate of $\epsilon \leq 1\,\text{mHa}$ which is $N_s \approx \mathcal{O}(10^2)$ for QPD and $N_s \approx \mathcal{O}(10)$ for sparse QPD. In Fig.\,\ref{fig:density_comparison}b, we also show (dashed lines) power-law $y = Ae^{-\alpha}$ fits to analyze the scaling of QPD and the different variants of sparse QPD. 

In the original work on QPD\,\cite{stroschein2026ground}, numerical experiments show Heisenberg scaling (i.e., $\alpha=1$) for both single- and multiple-eigenvalue estimation, provided that the signal length is sufficiently long for the given spectral density of the frequencies. In the analysis presented here, we did not explicitly test for Heisenberg scaling in our data. Instead, we allowed the exponent in our fitting model to vary freely, obtaining values closer to the dense-signal limit defined in Ref.\,\cite{dutkiewicz2022heisenberg}.

To better understand this result, we followed Ref.\,\cite{kimura2026fisher} and derived a Cramér–Rao lower bound (CRLB, see \ref{app:crlb}) for both the QPD algorithm, using the same shot schedule as in our numerical simulations, and sparse QPD, considering a number of shots independent of $T_{\max}$. In the first case, we observe that, under the rule $M_s = F\sqrt{N_s \text{ln}(N_s)}$, the CRLB allows for values of $\alpha$ up to $\alpha=0.7$, which is slightly lower than the values we observe numerically. We attribute this discrepancy to the fact that the CRLB does not account for the system knowledge that can be incorporated into an algorithm or for the specific algorithmic steps involved. Instead, it relies solely on the information-theoretic content of the measured data. In this respect, QPD exploits substantial domain knowledge, as the filter center $\omega^*$, the filter width $W_f$, and the sampling rate $W_s$ are either estimated classically or depend on the quality of the initial state used to compute $C(t)$.

These results are consistent with the recent work of Ref.\,\cite{leimkuhler2026quantumalgorithmweaklycorrelated}, which rigorously demonstrates that QPD can operate in the Heisenberg-limited regime. Together, these findings reconcile our results with those of Ref.\,\cite{ding2024esprit}, which shows that uniform sampling designs are incompatible with Heisenberg scaling.

Turning to the results for sparse sampling, we observe slightly higher values of $\alpha$ for both uniform and Gaussian sampling densities. This is consistent with the results presented in \ref{app:crlb}, where we derive Heisenberg scaling in this regime. This aspect requires further investigations, with a focus on the interplay between different shot and sampling schedules for quantum subspace methods, both with and without sparse sampling.

Figure\,\ref{fig:density_comparison}a shows the error relative to the ground-state energy as a function of the maximum signal length, $T_{\max}$. First, let us compare the results obtained with standard QPD with those obtained with sparse QPD for a uniform sampling density (yellow squares). The error introduced by signal recovery from sparse sampling degrades the performance compared to standard QPD. This result can be explained in terms of the additional errors introduced by matrix completion and rank projection. In this case, the former should be more significant than the latter, since the squared overlap with the ground state is very high for this system ($\approx 0.99$; see also Sec.\,\ref{sec:overlaps}), and further projection onto the ground-state-spanned subspace is already naturally incorporated into the QPD framework when projecting the GEP onto the $m$ leading eigenvectors of the $B$ matrix. 

We now turn to studying the effects of the different sampling densities. As we can see in Fig.\,\ref{fig:density_comparison}a, even though the exponential bias (pink solid line) is capable of minimizing the performance loss introduced in terms of $T_{max}$ scaling, there is no real advantage in terms of circuit depth around our target accuracy of $\epsilon \approx 1 \text{mHa}$. Our numerical experiments showed that the accuracy of the protocol becomes quite unstable with respect to the choice of the exponential decay parameter $\beta$. Moreover, when considering the sampling scheme obtained from a truncated Gaussian, we are able to achieve chemical accuracy with shorter circuits (here we have set $\tau = 0.5$ for these calculations) as compared to uniform and exponentially biased sampling. Since this reduction in circuit depth does not come with a substantial performance loss in terms of $T_{runtime}$  we use this sampling scheme as the default from here on. We also note that previous results using a truncated Gaussian as sampling distribution support this choice and inspired us during our development\,\cite{li2023note,ding2024quantum}.

To complete our initial analysis of the effect of sparse sampling on the computational cost and accuracy of QPD, we evaluated the number of points $N_s$ and the number of shots $N_{\text{shots}}$ per point as a function of signal length. As anticipated, the first observation is an exponential reduction in sampling cost. As shown in Fig.\,\ref{fig:density_comparison}c, the number of measured points is slightly different for Gaussian sampling. However, this discrepancy is solely due to the fact that we used two different multiplicative constants $C_{cs}$ (see Sec.\,\ref{sec:methods}) for the different sampling densities, specifically, $C_{\text{gauss}} = 3$ and $C_{\text{non-gauss}} = 2$. Finally, we look at Fig.\,\ref{fig:density_comparison}d, where we report the number of shots per sample used in this work. Here, we see that sparse QPD is able to maintain the  stability of QPD with respect to shot noise. In particular, the only difference we observe between sparse QPD and standard QPD is that, when using a sampling distribution that results in an effective $T_{max}$ lower than the actual value corresponding to the full grid, we have a slightly larger number of shots per sample. 

In practice, this increase is negligible compared to other quantum subspace methods in the literature, where the number of shots per sample can range from $N_{shots} = 10^3$ to $10^6$~\cite{shen2025estimating, lee2024sampling, klymko2022real}. By comparison, both QPD and sparse QPD operate in the regime of $N_{shots} \approx 10$. Even when accounting for sophisticated techniques to reduce the sampling cost, such as those developed in Ref.\,\cite{lee2025efficient}, the number of shots per sample required remains larger by several orders of magnitude.

\begin{figure}[h!]
        \centering
        \includegraphics[scale=0.85]{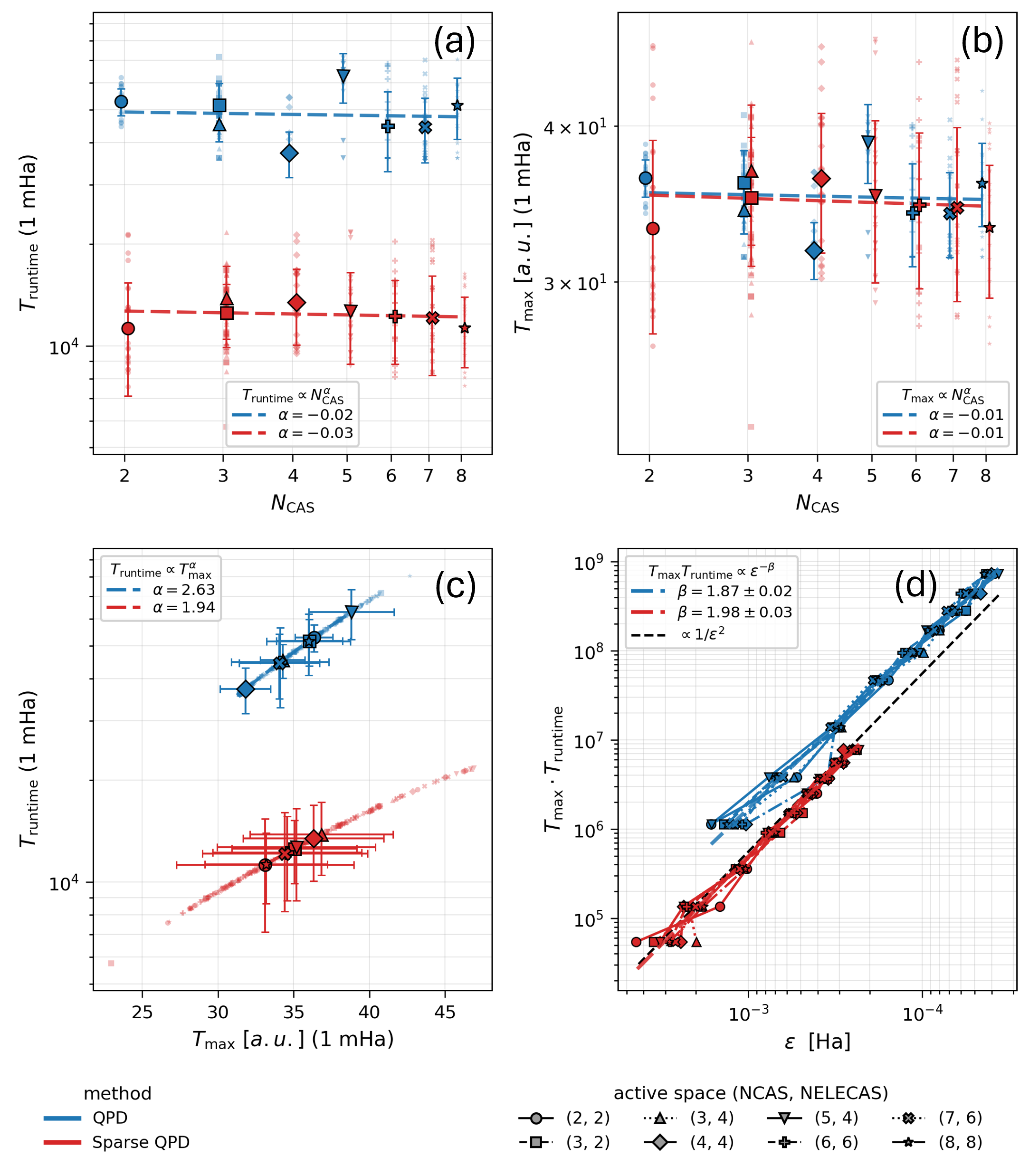}
        \caption{Scaling of sparse QPD with increasing active space size. (a) Minimal $T_{runtime}$ and (b) minimal $T_{max}$ required to achieve 1 mHa accuracy as a function of the active space sizes considered. Faint dots correspond to the values for each of the ARG-TRP geometries considered, while thick markers represent the median values for each active space and error bars show the corresponding interquartile ranges. Dashed lines correspond to power-law fits, $T_{runtime}(1\,\mathrm{mHa}) = b N_{CAS}^{\alpha}$. (c) Minimal runtime $T_{runtime}$ as a function of the minimal $T_{max}$ required to achieve chemical accuracy. The same plotting conventions as in panels (a) and (b) are adopted. (d) Empirical verification of the Cramér--Rao lower bound for QPD and sparse QPD\,\cite{kimura2026fisher}.}
        \label{fig:scaling_active_space}
    \end{figure}

Overall sparse QPD is capable of reducing the computational cost by approximately one order of magnitude, without requiring a substantial increase in the circuit depth needed to achieve chemical accuracy. These conclusions are based on a numerical analysis using a small active space of up to 3 spatial orbitals, because we can compare these results with the scaling observed in the presence of Trotter error and hardware noise. Nevertheless, from an algorithmic standpoint, it is important to understand how our method performs as the system size increases. We first focus on the upper panels (Fig.\,\ref{fig:scaling_active_space}a-b), where we report the scaling of $T_{runtime}$ ($T_{max}$) required to achieve $\epsilon = 1\,\text{mHa}$ as the system size increases. We observe two main features: (i) both $T_{runtime}$ and $T_{max}$ are independent of the system size for the systems studied, and (ii) with $\tau = 0.5$ used in these calculations, sparse QPD is able to obtain chemically accurate energies at the same circuit depth while requiring much smaller values of $T_{runtime}$. Let us first discuss the first observation. In particular, we can interpret this result in light of Eq.\,\eqref{eq:error_bound}. As mentioned earlier, several factors affect the accuracy of QPD. The computational cost and $T_{max}$ are essentially independent of the size of the active space because the complexity of the system enters (sparse) QPD only through the complexity of $C(t)$. In particular, the error introduced by the algorithm depends on the number of frequencies lying outside $B_f$ (the first term in the numerator of Eq.\,\eqref{eq:error_bound}), and the error introduced by shot noise (the second term in the numerator). It is therefore natural to ask whether these factors depend on the size of the active space. 

Indeed the first term depends solely on the width of the filter window, $W_f$, which must, in any case, satisfy $W_f > \frac{\pi\,\text{n}_{freqs}}{T}$, where $\text{n}_{freqs}$ is the number of frequencies within $B_f$~\cite{stroschein2026ground}. At the same time, the number of frequencies inside (or outside) $B_f$ depends, to a first approximation, on $|\langle \varphi_0|\Phi\rangle|^2$. In general, this quantity depends on the system size and, in particular, decays exponentially as the dimension of the Hilbert space increases~\cite{chan2012low}. For weakly (or moderately) correlated systems, however, it is possible to prepare states efficiently while maintaining a high overlap with the ground state~\cite{fomichev2024initial,erakovic2025high}. In Appendix\,\ref{sec:overlaps}, we discuss how this is the case for the system considered here, for active spaces of up to 70 spatial orbitals. Finally, the shot-noise contribution to the second term in the numerator of Eq.\,\eqref{eq:error_bound} is, in turn, determined solely by $T_{\max}$ through our sampling schedule. Consequently, neither $T_{\mathrm{runtime}}$ nor $T_{\max}$ exhibits system-size dependence.

Given the results of Fig.\,\ref{fig:scaling_active_space}, we can also infer that the signal recovery step does not add size dependent errors. In this case, one must use caution in accepting this statement, as the error that is clearly neglected by the projection of the Hankel matrix on a rank-reduced subspace is related to the residual weight 1 - $|\langle \varphi_0|\Phi\rangle|^2$ (see Sec.\,\ref{sec:cs}). Fortunately, for this quantity the same arguments based on Ref.\,\cite{fomichev2024initial,erakovic2025high} apply. 

Panel c reports the scaling of $T_{runtime}$ required to achieve chemical accuracy as a function of the corresponding $T_{max}$. The signal recovery preprocessing not only decreases the total computational cost without increasing the required circuit depth, but also reduces the scaling of $T_{runtime}(T_{max})$, modeled here with a power-law fit, from $T_{runtime} \propto \mathcal{O}(T_{max}^{2.63})$ to $T_{runtime} \propto \mathcal{O}(T_{max}^{1.94})$. Note that this result holds as long as we take into account the dependence of both the sampling cost and the completion error on $p_0$. We plan to explore this additional dependence in a future study. Turning to panel d of Fig.\,\ref{fig:scaling_active_space}, we discuss our finding in the context of Ref.\,\cite{kimura2026fisher}. There, the authors study the cost of phase estimation through a careful analysis of the Fisher information matrix, as initially introduced in Ref.\,\cite{dutkiewicz2022heisenberg}. For the purposes of our discussion, we follow the author's inequality relating $T_{runtime}$, $T_{max}$, and $\epsilon$. In particular, they find that phase-estimation approaches based on the Hadamard test circuit must satisfy $T_{runtime}T_{max}\epsilon^2 \geq \mathcal{O}(\frac{1}{p_0})$. According to both the fit and the measured data points, for all the structures and active spaces considered in this study, sparse QPD follows very closely this bound, with a fitted exponent of $\beta \approx 1.98 \,\pm 0.03$.

In this section, we have seen the effect of sparse sampling on QPD and how it relates to the original formulation of Ref.\,\cite{stroschein2026ground}. In the next section, we move on to study the effects of other sources of error on the measured $C(t)$ when adopting (or not) a signal recovery preprocessing step.

\subsection{Energy estimates on a quantum hardware}
\label{quantinuum_results}

We now report the experimental results of the estimation of the molecular ground state energy of the ARG-TRP pairs for a small active space, (2o,2e), within chemical accuracy.

To this end, we used the Quantinuum Helios quantum device to execute quantum circuits.
Quantinuum Helios~\cite{ransford2026} is a 98-qubit trapped-ion processor built on the quantum charge-coupled device (QCCD) architecture,
with all-to-all connectivity achieved by ion transfer between operational zones via an X-shaped junction.
The native gates are arbitrary-angle single-qubit rotations, virtual
$R_Z$ rotations, and an arbitrary-angle M{\o}lmer--S{\o}rensen-type
entangler $R_{ZZ}(\theta)$.
Reported average
infidelities are $2.5(1)\times10^{-5}$ for single-qubit gates,
$7.9(2)\times10^{-4}$ for two-qubit gates, and $4.8(6)\times10^{-4}$
for state preparation and measurement~\cite{ransford2026}.
Programs are compiled to the HUGR intermediate representation~\cite{koch2025}.
The companion emulator, Helios-1E, running on a GPU backend,
executes the same HUGR programs.
In our case, that is circuits of up to 32-qubits, it is a statevector simulator.
The hardware and the emulator are reached through the Quantinuum Nexus cloud platform; compiled programs are uploaded and executed as asynchronous jobs.

\subsubsection*{Single-point calculation on a trapped ion quantum hardware}

\begin{figure}[h!]
        \centering
        \includegraphics[width=\linewidth]{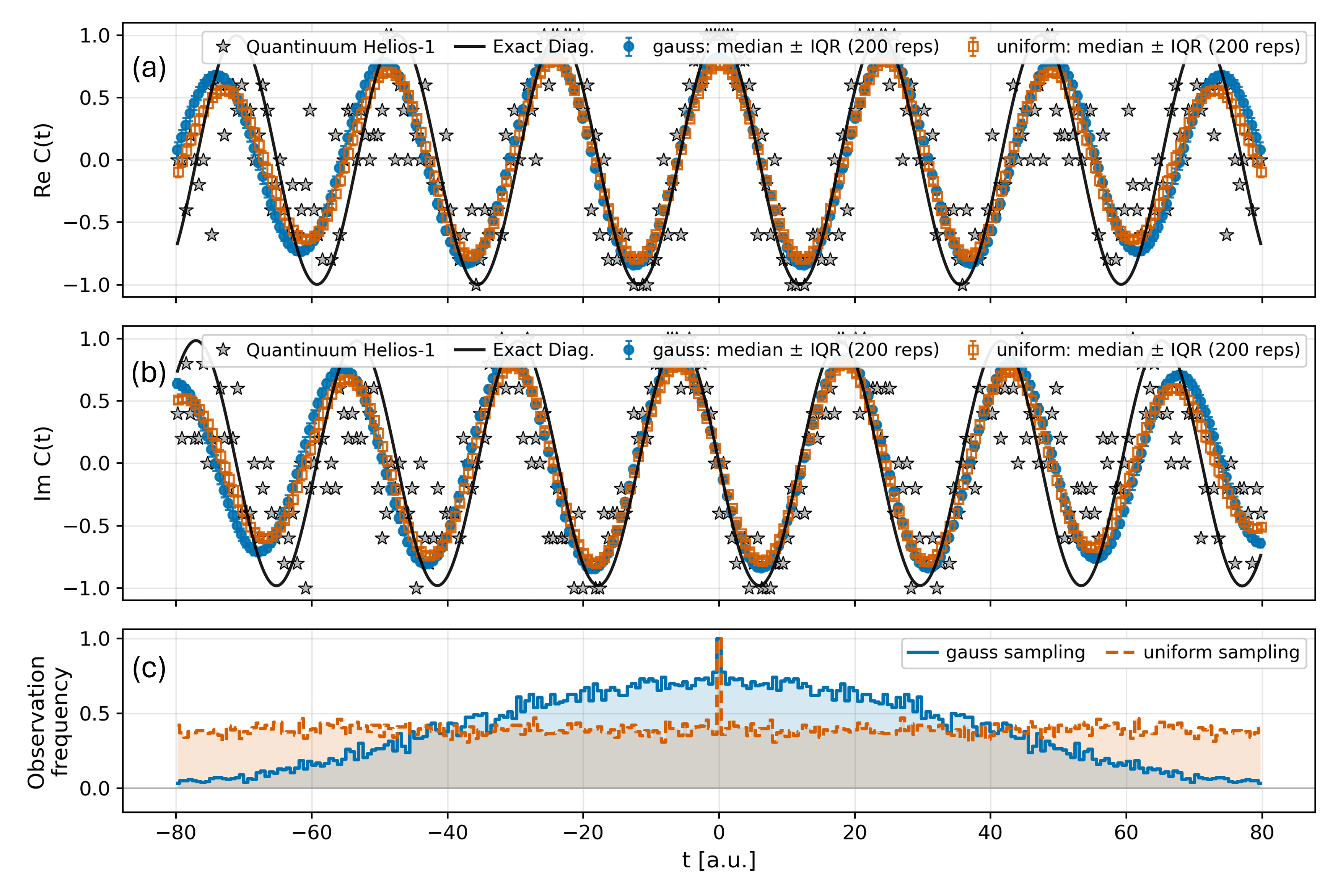}
        \caption{(a, b) Recovering autocorrelation functions $C(t)$ from samples measured on a quantum computer. Stars show raw measurements from the Quantinuum Helios-1 machine, blue dots (orange squares) correspond to the median values obtained recovering the signal from 200 different subsets (compression factor of $\approx 0.39$) of the full dataset. Error bars indicate the interquartile range. (c) Comparison between the different densities studied (see Eq.\,\eqref{eq:densities}). Note that the spike corresponding to the value of $t=0$ is due to the fact we always include this value as, by definition, $C(0) = 1$.}
        \label{fig:signal_recovery}
\end{figure}

We first consider the measurements collected on the Helios machine, reported in Fig.\,\ref{fig:signal_recovery} (stars). At early times, the measurements closely follow the exact evolution (black solid line) and progressively depart from it as $t$ increases. To characterize the statistical behavior of the recovery with the HSNLD algorithm, we formed 200 different subsets with a compression factor of 0.39 (i.e., we used 50 data points for the recovery and $N_s = 128$). In the figure, we report the median values of the recovered signal using uniform (Gaussian) sampling in orange (blue). We observe that the median of the recovered signal deviates from the exact diagonalization also because it shows a decay over time. This is due to the fact that, by allowing for complex amplitudes, the recovery captures the decay due to decoherence in the quantum hardware. In Fig.\,\ref{fig:signal_recovery}c, we report the frequency of each sample over the 200 repetitions according to the chosen sampling density. For the Gaussian sampling, we set $\tau = 0.85$, resulting in an effective $T_{max}^{\mathrm{Gaussian}} = 67.83\,\mathrm{a.u.}$. For both the uniform sparse QPD and the standard QPD calculation, we have $T_{max} = 79.80\,\mathrm{a.u.}$.

\begin{figure}[h!]
        \centering
        \includegraphics[width=\linewidth]{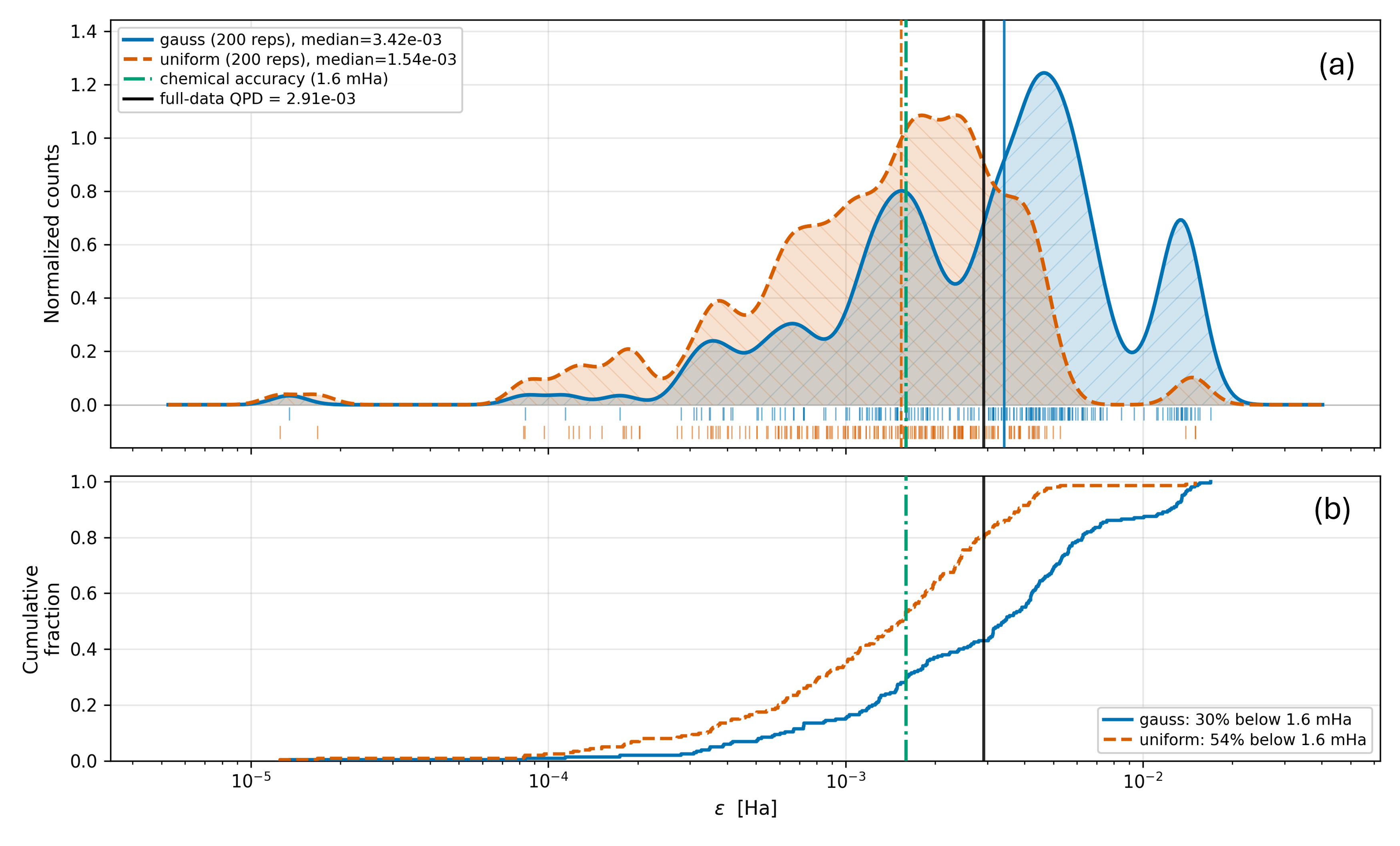}
        \caption{QPD and sparse QPD energy estimates from the data shown in Fig.\,\ref{fig:signal_recovery}. (a) Distribution of sparse QPD ground-state errors for different initial measurement guesses (i.e. different $\textbf{f}$, as specified in Eq.\,\eqref{eq:obs}, obtained from the full measurement set $\tilde{\textbf{x}}$). Bars show the empirical distributions, with solid lines representing Gaussian kernel density estimates (KDEs) of the corresponding distributions, as implemented in SciPy\,\cite{virtanen2020scipy}. The bandwidth factor is set to 0.35. (b) Cumulative fraction of the empirical distribution as a function of the ground-state energy error.}
        \label{fig:energy_distribution}
\end{figure}

From the signal distribution shown in Fig.\,\ref{fig:signal_recovery}, we characterize the corresponding error distribution obtained with QPD for the ground-state energy error. This distribution is displayed in Fig.\,\ref{fig:energy_distribution}. The error estimate of the full QPD calculation is $\epsilon_{QPD} \approx 2.9\,\text{mHa}$. We can therefore compare this value with the error estimates obtained using Gaussian sparse QPD, $\epsilon_{G,SQPD} \approx 3.42\,\pm\,4.14\,\text{mHa}$, and uniform sparse QPD, $\epsilon_{U,SQPD} \approx 1.54\,\pm\,1.77\,\text{mHa}$, where we use the interquartile range as a measure of uncertainty. The recovery from different subsets of measurements is characterized by a non-negligible degree of multimodality, which is attributed to the different minima found by the HSNLD algorithm. Statistically speaking, given the associated uncertainties, it is not possible to conclude that any of the three methods provides more accurate results than the others. Due to constraints on machine access, we could perform only one measurement run for the autocorrelation function. Consequently, we could not accumulate sufficient statistics to estimate the uncertainty associated with the error obtained using QPD on the full dataset. Similarly, we are aware that the sampling of the signal-recovery and associated-error distributions is biased by the fact that we use a single dataset to generate the subsets from which the signals are recovered.

Nevertheless, given these limitations in our analysis, we are able to show that QPD and sparse QPD can provide chemically accurate energy estimates (i.e. the value of $\epsilon = 1.6 \text{mHa}$ is within the experimental error bars) implementing a controlled evolution using the Trotterized time evolution circuit as detailed in Appendix\,\ref{sec:trotter}. Our implementation of time evolution enables us to achieve maximal circuit depths of 448 two qubit gates and total gate count of 1221 on a 3-qubits circuit register (2 qubits encoding the wavefunction and one ancillary qubit). 
We compared our results with recent results in literature that report an experimental implementation of phase estimation protocols to estimate ground state energies~\cite{o2016scalable,blunt2023statistical, yamamoto2024demonstrating, yamamoto2026quantum, nutzel2026ground, kanno2025tensor, sorathia2025quantum}.
Ref.\,\cite{o2016scalable} presents the first experimental implementation of a phase estimation algorithm for quantum chemistry. The authors report ground-state energies for the $\text{H}_2$ molecule in the minimal basis set, obtained using a modified version of Kitaev's iterative phase estimation algorithm\,\cite{kitaev1995quantum}. The phase estimation experiment uses three qubits: one ancillary qubit to implement a Trotterized evolution and two qubits to encode the molecular wavefunction. Due to circuit-depth constraints, the time evolution is approximated using a single Trotter step. Through careful control of the gate pulses, the authors achieve dissociation energy errors of approximately $10\,\text{mHa}$. More recent experiments have improved upon this result, benefiting from advances in quantum hardware. In particular, Blunt \textit{et al.}\,\cite{blunt2023statistical} report an experiment combining the algorithms proposed in Refs.\,\cite{ding2023even,wan2022randomized}. Although they demonstrate experiments involving larger circuits (up to seven qubits), their implementation is not scalable, as it relies on variational simulation techniques\,\cite{li2017efficient} (similar experiments on even larger systems have been demonstrated in Ref.\,\cite{kanno2025tensor, sorathia2025quantum}). For the experiments involving Trotterized evolution, they report a ground-state energy error of $\epsilon = 8\,\text{mHa}$ for the $\text{H}_2$ molecule, mapped onto an effective single-qubit Hamiltonian. The corresponding circuits contain up to 100 two-qubit gates. 

We now turn to more recent experiments performed on similar trapped-ion devices as the one used in this work~\cite{pino2021demonstration, moses2023race}. In Ref.\,\cite{yamamoto2024demonstrating}, the authors demonstrate the integration of an error-detection protocol\,\cite{self2024protecting} with a Bayesian quantum phase estimation (QPE) algorithm\,\cite{wiebe2016efficient}. Using a first-order Trotter approximation for the time evolution, the encoded circuits contain up to 920 two-qubit gates. The authors achieve a ground-state energy error of $\epsilon = 6\,\text{mHa}$, again for the $\text{H}_2$ molecule, represented by an effective two-qubit Hamiltonian. Building on this result, a subsequent work demonstrates the integration of an error-correction protocol\,\cite{steane1996multiple} with the phase estimation algorithm of Ref.\,\cite{svore2013faster}, using a single-logical-qubit representation of the $\text{H}_2$ molecule. Their experiment involves up to 16 physical qubits and reports conditional and unconditional two-qubit gate counts of 7202 and 1585, respectively, together with a ground-state energy error of $\epsilon = 1.00\,\pm\,13\,\text{mHa}$.

Finally, we discuss the only experiment to date that goes beyond the $\text{H}_2$ molecule: the recent work of Ref.\,\cite{nutzel2026ground}, which reports a ground-state energy estimate for the $\text{H}_3^+$ molecule encoded in a six-qubit space. The circuits implement adiabatic state preparation using a randomized algorithm\,\cite{granet2024hamiltonian,granet2025practicality}, with an average of 993 two-qubit gates and a ground-state energy error estimate of $\epsilon = 25.5\,\pm\,5.3\,\text{mHa}$. 

Even though these results illustrate how the joint advancement of hardware quality, algorithmic efficiency, and compilation optimization allows us to push the boundaries of what is currently achievable on a programmable quantum device, they provide less insight into the behavior of sparse QPD in the presence of hardware noise and evolution error. For this reason, in the next section, we numerically study the scaling properties of QPD and sparse QPD using a noise model that has been validated against real quantum hardware (see Appendix\,\ref{app:helios} for further details).

\subsubsection*{Performance under Trotter error and Hardware noise}

\begin{figure}[h!]
        \centering
        \includegraphics[scale=0.85]{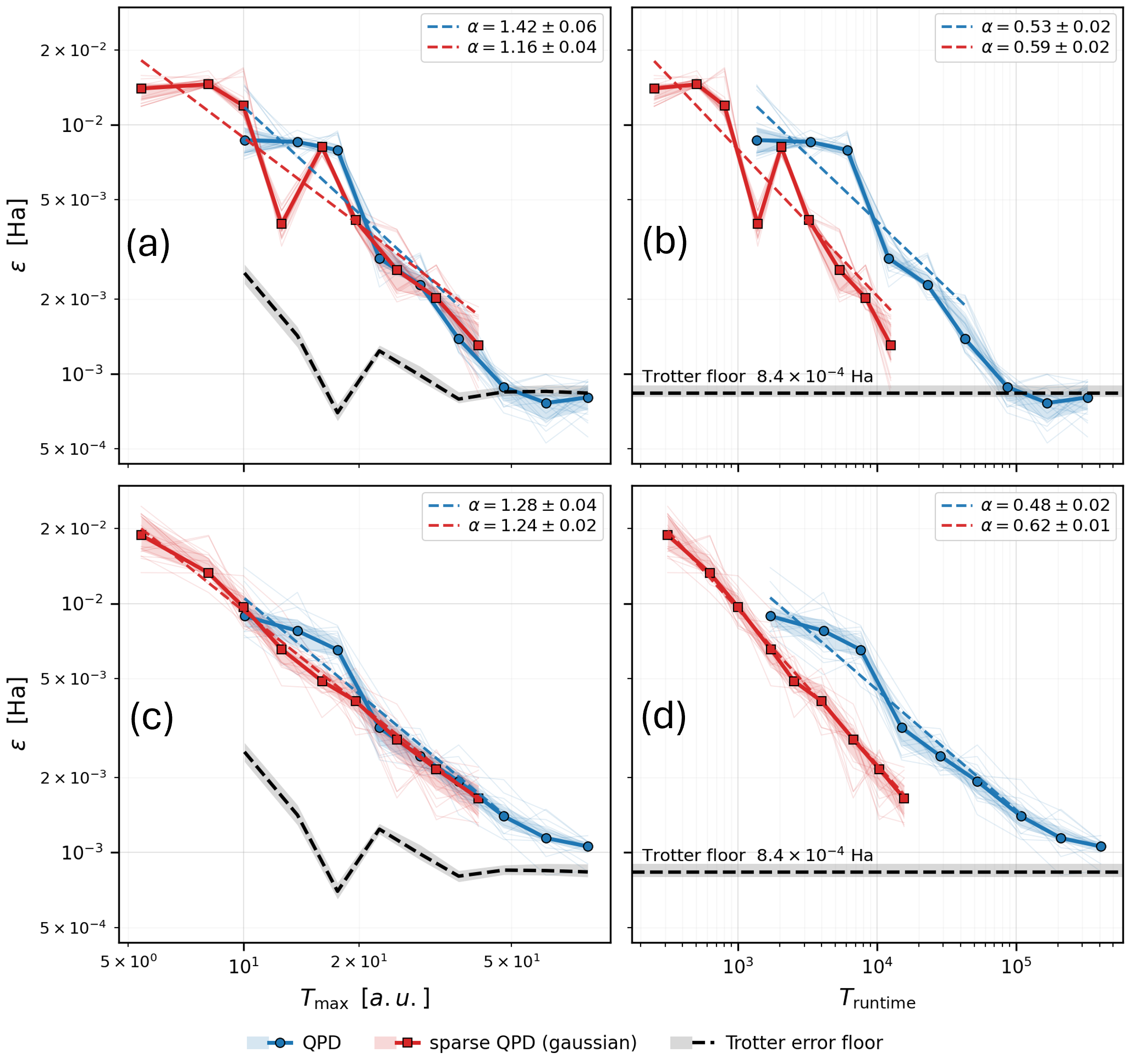}
        \caption{Scaling analysis of (sparse) QPD with Trotter error only (a, b) and Trotter error plus hardware noise (c, d) for an active space of (2o, 2e). Ground state energy error as a function of the maximal evolution time $T_{max}$ (a, c) and the total runtime $T_{runtime}$ (b, d). Faint solid lines are obtained by averaging over 10 realizations of a (sparse) QPD calculation for a given value of $T_{max}$ and correspond to the different ARG-TRP geometries, as explained in the computational details section. Thick solid lines show the median values computed from the estimates for the different geometries, while the shaded areas correspond to the interquartile ranges. Dashed lines show power-law fits, $\epsilon = a x^{-\alpha}$, where $x = \{T_{max}, T_{runtime}\}$. The Trotter floor is computed by evaluating the error obtained with a QPD calculation using data from a noiseless (i.e., infinite-shot) execution of the time-evolution circuit. The grey shaded area represents the interquartile range of the Trotter floor across the different ARG-TRP structures.}
        \label{fig:noise_scaling_small_active_space}
    \end{figure}

In Fig.\,\ref{fig:noise_scaling_small_active_space} we report the same scaling plots that we have discussed in Fig.\,\ref{fig:density_comparison} obtained for the ARG-TRP geometries but now including the effect of the Trotter evolution (Fig.\,\ref{fig:noise_scaling_small_active_space}a-b) and hardware noise (Fig.\,\ref{fig:noise_scaling_small_active_space}c-d); these results are obtained on a subset of 18 out of 23 geometries due to the additional computational overhead of simulating hardware noise. First, we focus on a smaller active space of (2o, 2e). We start considering the results in Fig.\,\ref{fig:noise_scaling_small_active_space}a. Looking at the panel showing the error as a function of $T_{max}$, we see that the $T_{max}$ corresponding to $1\,\text{mHa}$ is increased in the presence of a systematic error due to Trotterization, rising from approximately $T_{max} = 35 a.u.$ (dots in Fig.\,\ref{fig:scaling_active_space}b) to approximately $T_{max}(1\,\text{mHa}) = 45–50\,\, a.u.$. As for the scaling of the error as a function of $T_{runtime}$, the advantage of sparse QPD is evident in this case as well. The key difference compared with the previous results, where the sole source of noise was finite sampling, is that the error does not decrease indefinitely as computational resources increase. Instead, as expected, it stabilizes at a value determined by the Trotterization error, which becomes the dominant source of error.

Including hardware decoherence (Fig.\,\,\ref{fig:noise_scaling_small_active_space}c-d), we observe that introducing this additional source of error further increases the $T_{\max}$ required to reach $\epsilon=1\,\text{mHa}$. In particular, the blue curve reaches a plateau more slowly, and the value at which it stabilizes is slightly higher than the Trotter floor. This behavior can be explained by the additional bias introduced by executing the simulated circuit on actual quantum hardware. Indeed, not only does the Trotterized evolution differ from the exact evolution, but hardware noise also affects the circuit execution increasing the \textit{effective} noise floor. These results are also in line with our single-point calculation on the quantum hardware as the uncertainties that we observe over this much larger set of calculations confirm that both QPD and sparse QPD are able to achieve $\epsilon \approx 1\,\text{mHa}$ for $T_{max} \approx 65\,a.u.$\,.

\begin{figure}[h!]
    \centering
    \includegraphics[scale=0.85]{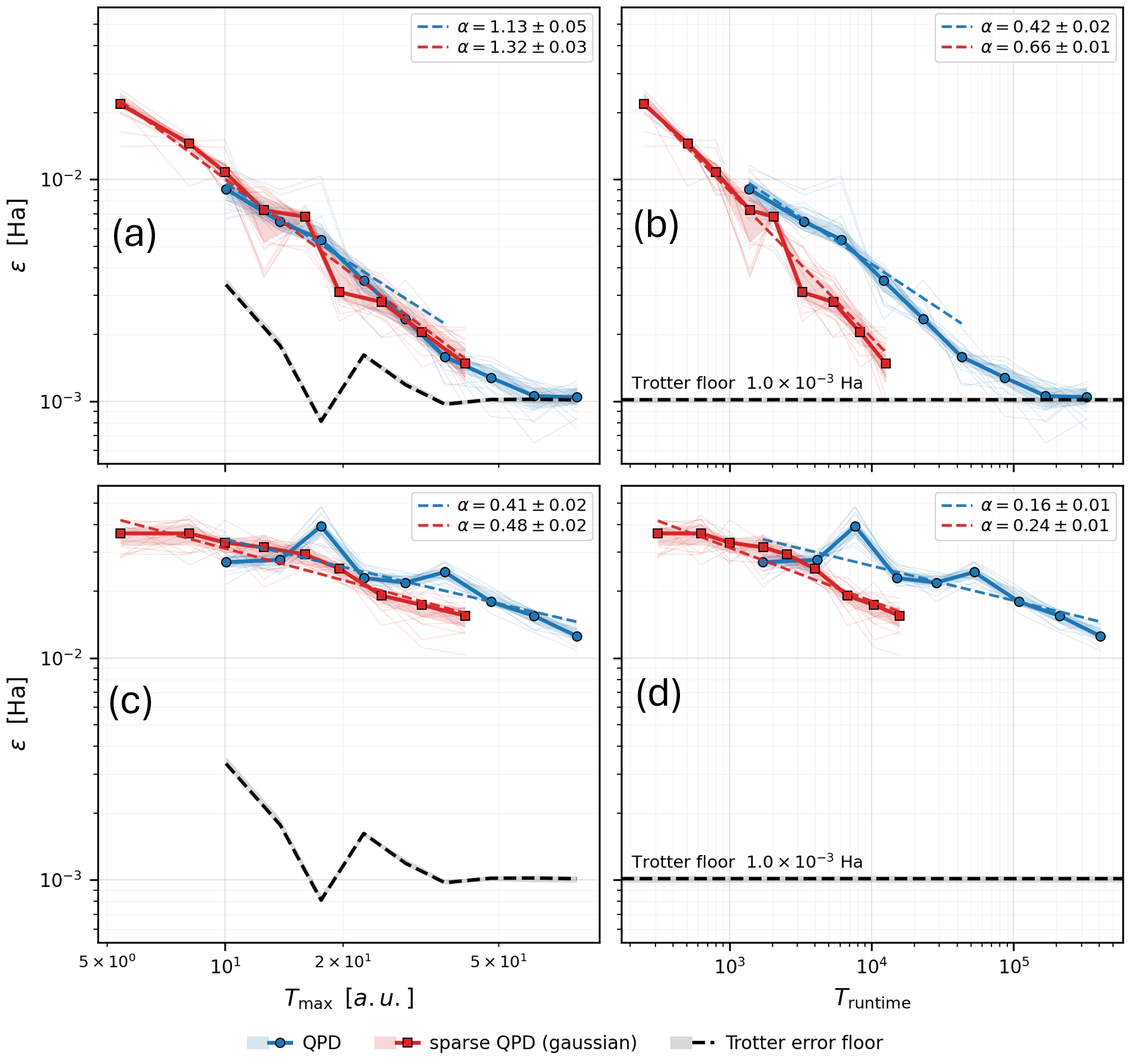}
    \caption{Scaling analysis of (sparse) QPD with Trotter error only (a, b) and Trotter error plus hardware noise (c, d) for an active space of (2e, 3o). Ground state energy error as a function of the maximal evolution time $T_{max}$ (a, c) and the total runtime $T_{runtime}$ (b, d). Faint solid lines are obtained by averaging over 10 realizations of a (sparse) QPD calculation for a given value of $T_{max}$ and correspond to the different ARG-TRP geometries, as explained in the computational details section. Thick solid lines show the median values computed from the estimates for the different geometries, while the shaded areas correspond to the interquartile ranges. Dashed lines show power-law fits, $\epsilon = a x^{-\alpha}$, where $x = \{T_{max}, T_{runtime}\}$. The Trotter floor is computed by evaluating the error obtained with a QPD calculation using data from a noiseless (i.e., infinite-shot) execution of the time-evolution circuit. The grey shaded area represents the interquartile range of the Trotter floor across the different ARG-TRP structures.}
    \label{fig:noise_scaling_large_active_space}
\end{figure}

To conclude, we increase the system size (i.e., the number of operations) to assess the behavior of (sparse) QPD in a regime where the noise exceeds a tolerable threshold. Fig.\,\ref{fig:noise_scaling_large_active_space} presents the same scaling analysis as Fig.\,\ref{fig:noise_scaling_small_active_space}, but for an active space of (3o, 2e), for which the resulting compiled quantum circuits contain five qubits. When considering only the effect of Trotter error, the conclusions are identical to those drawn from Fig.\,\ref{fig:noise_scaling_small_active_space}a-b. However, when hardware noise is included, the results differ substantially. The noise accumulated due to the increased number of gates is so significant that, even after noise suppression through low-rank projection and prolate filtering, the estimated error is an order of magnitude larger than in the previous case. A different behavior is also observed in the scaling exponent. As noted for the smaller active space, introducing hardware decoherence reduces the power-law exponent, and this effect becomes more pronounced as the hardware noise strength increases. This is because as the hardware noise strength increases, progressively less information can be extracted by increasing $T_{max}$, and hence $T_{runtime}$. Evidence for this is that the $\epsilon$ values of Fig.\,\ref{fig:noise_scaling_large_active_space}c-d and Fig.\,\ref{fig:noise_scaling_small_active_space}c-d are comparable at low $T_{max}$ (i.e. in the short circuit regime) and that the effect of increasing the hardware noise becomes more evident for larger values of $T_{max}$. Ultimately, one would expect the error to plateau at sufficiently large values of $T_{max}$ and $T_{runtime}$, as observed in the previous figure. This plateau would reflect the combined effect of the Trotter error floor and an additional bias arising from other sources of noise.

\section{Conclusions}

We presented a variant of QPD, denoted sparse QPD, in which the autocorrelation function is measured only on a random subset of the QPD time grid, while the remaining samples are reconstructed classically using Hankel matrix completion\,\cite{Cai2025}. The method relies on a property of weakly correlated systems: when the initial state has a dominant overlap with the ground state, the Hankel matrix constructed from samples of $C(t)$ is close to a low-rank Hankel matrix. Consequently, the number of distinct circuits that need to be executed scales as $\mathcal{O}(\log(N_s))$, rather than $\mathcal{O}(N_s)$. The completed signal is processed using the original QPD algorithm, preserving the error controllability of the original framework, provided that the completion error is negligible when compared with shot noise. 

With the long-term goal of investigating how quantum computing can accelerate~\cite{castaldo2025heisenberg} and improve quantum chemical workflows~\cite{husistein2026new} (such as the training of machine-learned interatomic potentials~\cite{behler2007generalized, bartok2017machine,husistein2026new} for chemical reactions and conformational sampling), we selected 23 representative structures of a ARG-TRP pair from a molecular dynamics trajectory and evaluated the algorithmic performance of sparse QPD on these structures. Compared with standard QPD, sparse QPD reduces the computational cost of ground-state phase estimation up to an order of magnitude.
Furthermore, we experimentally demonstrated the entire workflow on a trapped-ion quantum device. It comprised the incorporation of the solvating environment into the molecular Hamiltonian through electrostatic embedding, its representation on the quantum computer, the explicit compilation of the Trotter–Suzuki evolution circuits, and, finally, matrix completion and the solution of the generalized eigenvalue problem on a classical computer. Ground-state energies consistent with chemical accuracy within statistical uncertainty were estimated, without employing any error mitigation, detection, or correction protocols, using circuits containing up to 448 two-qubit gates and 1,221 gates in total.

Several aspects of this work require further investigation. First, our claims are based on numerical studies and extending the formal results of Ref.\,\cite{Cai2025} to different sampling densities and levels of shot noise would provide further insights into the proposed method. Second, the completion rank was fixed to one throughout the study, reflecting the high overlap between the Hartree--Fock determinant and the ground state. Although the effect of the Hankel matrix rank on the sampling cost has been characterized by Cai \textit{et al.}\,\cite{Cai2025}, making this parameter controllable, we plan to investigate this aspect in greater detail in future work. Third, we derived Cramer--Rao lower bounds for both QPD and sparse QPD, complementing the non-asymptotic error analysis reported in Ref.\,\cite{stroschein2026ground} and the end-to-end characterization presented in Ref.\,\cite{leimkuhler2026quantumalgorithmweaklycorrelated}. How these results can be harmonized with the already existing scaling analysis will be subject of a future work. 

Finally, the work of Ref.\,\cite{leimkuhler2026quantumalgorithmweaklycorrelated} specifically focuses on the weakly correlated regime, showing that the overall scaling of QPD with system size is asymptotically $\mathcal{O}(n^{4.8})$, where $n$ denotes the number of spin-orbitals.  Together with these scaling results, the speedup demonstrated in this work suggests that standard quantum chemistry, which is typically dominated by weakly correlated electronic structures, can be tackled by future routine quantum computations, with an advantage over the best classical approaches.

\begin{acknowledgments}
\section*{Author contributions}

D.C. devised the sparse QPD algorithm, and wrote the sparse QPD software to extract the autocorrelation function from quantum measurements.
M.K. prepared the Hamiltonians, selected the molecular structures that were employed, as well as compiled QPD quantum ciruits.
M.M. ran circuits on the Quantinuum hardware, 
and performed the randomized benchmarking to compare against the noise model.
D.C. built the classical noise model based on publicly available information of the Helios hardware specifications.
S.F. performed MD simulations and performed the clustering analysis to extract representative structures for the different amino acid pairs.
G.C.S., K.L.-L., M.C., and M.R. planned and supervised the project and acquired the financial funding and compute. 
D.C., M.K., M.M., S.F. wrote the first version of the manuscript.
All authors reviewed the manuscript.

\section*{Statement on LLM use}
LLMs were not used to design the study or to formulate its research questions.
LLMs (Claude) were used in conversational mode to assist in code development and in the derivation of Sec.\,\ref{app:crlb}. 
Claude (Opus 5) agents were employed to automatize the computation performed on the Quantinuum Helios quantum computer (submit compute jobs using Quantinuum API, poll and fetch the results).
Open AI models were used for copy editing; the text and references were subsequently checked and approved by the authors.
The authors declare that the final statements and proofs are their own original work.

\section*{Funding}
The authors gratefully acknowledge financial support from XPRIZE through a Phase-2 finalist award.
M.C., M.M., M.R., D.C., M.K. acknowledge support from the Novo Nordisk Foundation (grant no. NNF20OC0059939 ‘Quantum for Life’). D.C. acknowledges support by Grant No. 233975 (SPEED-Q Swiss National Postdoctoral Fellowship).
M.R. gratefully acknowledges financial support from the Swiss National Science Foundation through Grant No. NF 200021\_219616.
S.F. and G.C.S. are supported by the Novo Nordisk Foundation, Grant number NNF22SA0081175, NNF Quantum Computing Programme. M.C., M.R., K.L.L. and G.C.S. gratefully acknowledge the NNF Synergy grant "Molecular Recognition from Quantum Computing" NNF25OC0105181. 
We acknowledge support from Danish e-infrastructure Consortium (DeiC) to provide the funding for the access to Quantinuum services (grant no. Q-ACCESS-H2-2025-02).

\section*{General Acknowledgments}
The authors thank Beatriz Piniello for assisting in the MD simulations, and Shobhit Sanjeev Chaturvedi for providing the MD simulation snapshot, which served as the input structure to our pipeline, for which we ran a calculation on the Quantinuum Helios computer. The authors thank Oskar Leimkuhler for useful comments on the manuscript.

\end{acknowledgments}

\appendix
\section{Low-cost Trotterization of the time evolution operator}\label{sec:trotter}
\subsection{Selection of Trotterization scheme.}\label{sec:trotter_scheme_det}

Multiple techniques exist to implement the
time evolution operator $U^{(t_k)}_H$ in Eq. \eqref{eq:htest_circ} on
quantum hardware~\cite{abrams1999quantum, childs2012hamiltonian, berry2015simulating, low2019hamiltonian}.
Trotterization is well suited to current non-fault-tolerant hardware with limited single- and two-qubit gate fidelities and has been employed in previous QPE experiments~\cite{graham2022multi, izsak2023quantum, tranter2025high, yamamoto2026quantum}.

Circuit depth, measured by the number two-qubit gates
is the main bottleneck for performing time evolution simulations on current, noisy quantum hardware.
Consequently, to make the calculation of molecular energies by QPD feasible on current hardware, one needs to carefully consider the selected Trotterization scheme.

For the tiny active spaces, where accurate QPD or
QPE calculations are feasible on current quantum devices,
the most suitable Trotterization scheme is expected to differ
from state-of-the-art product formula methods~\cite{gunther2026phase}
with more favorable asymptotic complexity.
This is because approaches based on randomized product formulas, e.g. Ref.~\cite{campbell2019random},
become increasingly effective, as the number of small-coefficient terms in the qubit
Hamiltonian increases.
This is the case for large active space sizes $n$, in particular, when a localized orbital basis
is employed.
The qubit Hamiltonians, which we consider for quantum hardware simulation in this work,
do not display this property. For example, the qubit Hamiltonian that was employed
for the quantum computation in section \ref{quantinuum_results} has a mere 8 terms.
We additionally considered recent numerical results~\cite{kronenberger2026trotter},
which reported larger Trotter errors for randomized product formulas~\cite{childs2019faster}
compared to a deterministic Trotter formula for systems with small active spaces.

We consequently decided to employ a deterministic Trotter formula,
as was also done for a recent demonstration of single-ancilla QPE
on quantum hardware~\cite{yamamoto2026quantum}.
For simplicity, we approximated the time evolution operator $U_{H}^{(t)}$ by
a first-order Trotter formula:
\begin{equation}
U^{(t)}_H = U_{\mathrm{Trot}}^{(t)} + O(t^2) \ ,
\end{equation}
with

\begin{equation}\label{eq:trotter_tevol}
U_{\mathrm{Trot}}^{(t)} = \prod_k \exp(- i t h_k P_k) \ ,
\end{equation}
given a qubit Hamiltonian $H_q$ of the form

\begin{equation}\label{eq:H_qubit}
H_q = \sum_k^{\Gamma} h_k P_k \ ,
\end{equation} 
where $P_k$ symbolize (non-identity) Pauli strings.

Apart from the order and type of product formula, 
other factors such as (i) term ordering, (ii) orbital basis,
and (iii) employed fermion-to-qubit mapping affect
circuit depth and Trotter errors for Trotterized Hamiltonian simulation.
We justify our choices for these below:

(i) the term ordering is mainly known to affect the Trotter error~\cite{tranter2019ordering},
but can also be exploited to reduce the circuit depth of a Trotter step by cancellation
of CNOT ladders (see Eq. \eqref{eq:cnot_ladder_circ})~\cite{li2022paulihedral}
between neighboring operators $\exp(- i t h_k P_k)$.
Finding heuristic ordering strategies, which produce comparatively low Trotter
errors with high consistency is challenging, although some reports
observed low errors to be produced by a reverse magnitude ordering
somewhat consistently~\cite{tranter2019ordering, kronenberger2026trotter}.
As such, we will instead choose term orderings that lead to shallow quantum
circuits by exploiting circuit compilation strategies for mutually commuting
sets of Pauli strings. Details are provided further below in Appendix \ref{sec:software_impl}.

(ii) The choice of orbital basis affects the circuit depth
of a Trotter step, since it affects the number of terms $\Gamma$
of the qubit Hamiltonian $H_q$ with non-negligible coefficients $h_k$~\cite{koridon2021orbital}.
Additionally, it affects the Hamiltonian 1-norm $\lambda$,
a quantity that correlates with upper bounds
on the Trotter energy error~\cite{childs2021theory, kronenberger2026trotter}.
Local orbital bases are known to minimize $\lambda$~\cite{koridon2021orbital}.
These bounds have been consistently found to be loose, however,
and suggest trends which are not observed in exact numerical experiments~\cite{mehendale2025estimating, kronenberger2026trotter}.
Our recent numerical results in Ref.~\cite{kronenberger2026trotter} show that
for small active spaces, localizing orbital transformations,
fail to consistently decrease the Hamiltonian 1-norm $\lambda$.
Additionally, we observed poor correlation between 1-norm
and Trotter error.
Consequently, we did not consider additional "orbital optimization" schemes,
and employed MP2 natural orbitals as the orbital basis in all experiments.

(iii) The fermion-to-qubit mapping affects the circuit depth of
a Trotter step, since it determines the average length of the Pauli strings
in the qubit Hamiltonian $H_q$.
For large system sizes, the Bravyi-Kitaev mapping~\cite{bravyi2002fermionic} is expected
to be preferred, as only $O(\log(n))$ qubit operations are required to implement
a fermionic operator on qubits, compared to the $O(n)$ operations
per fermionic operator following the Jordan-Wigner mapping~\cite{seeley2012bravyi}.
However, at the tiny system sizes for which we can realize simulations
on quantum hardware, the choice of fermion-to-qubit mapping
is not relevant yet.

Even after the selection of an appropriate Trotterization scheme
that facilitates short circuit depth Trotterized Hamiltonian simulation,
the simulation of multiple Trotter time evolution steps that would be required
for accurate sampling of the autocorrelation function by the Hadamard test
in Eq. \eqref{eq:htest_circ} for the smallest possible active spaces,
which involve 2 spatial orbitals, i.e. map to 4 qubits after the fermion-to-qubit mapping,
is prohibitively expensive for current hardware.

To further reduce the cost of Hamiltonian simulation, we
employ additional pre-processing of the qubit Hamiltonian by qubit tapering
techniques~\cite{bravyi2017tapering}:
By breaking the $S_z$ and particle number conservation symmetries
of the qubit Hamiltonian, a tapered qubit Hamiltonian $H_{q, \mathrm{tap}}$
can be obtained, which requires two fewer qubits, and has a slightly lower
number of terms, i.e. $\Gamma_{\mathrm{tap}} < \Gamma$.
We briefly point out that if one works with molecular structures or orbital
active spaces with point group symmetries, an additional qubit per symmetry operation can
be tapered from the qubit Hamiltonian~\cite{setia2020reducing}.
This has been exploited previously, e.g. in Ref.~\cite{yamamoto2026quantum}
where the minimal basis hydrogen molecule can then be simulated by an effective
single-qubit Hamiltonian.
However, such point group symmetries cannot be exploited for the orbital active spaces
of the ARG-TRP pair considered throughout this work, making their simulation more challenging.

\subsection{Synthesis of low-cost quantum circuits.}

As is apparent from Eq. \eqref{eq:trotter_tevol}, the implementation
of Trotterized Hamiltonian simulation requires implementation
of the unitary operators

\begin{equation}
 U_{\mathrm{Pg}} = \exp( i \theta_k P_k) \ ,   
\end{equation}
on a quantum computer. These operators $U_{\mathrm{Pg}}$ are commonly
referred to as "Pauli gadgets" in the literature.
It is well known that Pauli gadgets can be mapped to
quantum circuits efficiently as depicted in Eqs. \eqref{eq:upg_circuit}
and \eqref{eq:cnot_ladder_circ}~\cite{cowtan2019phase}:

\begin{equation}\label{eq:upg_circuit}
U_{\mathrm{Pg}} =
\Qcircuit @C=1.2em @R=1.0em {
   & \multigate{1}{V_1^\dagger} & \multigate{1}{V_2^\dagger} & \qw                & \multigate{1}{V_2} & \multigate{1}{V_1} & \qw \\
   & \ghost{V_1^\dagger}        & \ghost{V_2^\dagger}        & \gate{R_z(\theta_k)} & \ghost{V_2}        & \ghost{V_1}        & \qw
}
\end{equation}

with $V_1$ composed of single-qubit rotation gates, whereas the circuit $V_2$ contains the CNOT ladder.
\begin{equation}\label{eq:cnot_ladder_circ}
V_2^\dagger =
\Qcircuit @C=1.2em @R=1.0em {
    & \ctrl{1} & \qw      & \qw      & \qw       & \qw    \\
    & \targ    & \ctrl{1} & \qw      & \qw       & \qw    \\
    & \qw      & \targ    & \qw      & \qw       & \qw    \\
    & \vdots   & \vdots   & \vdots   & \vdots    & \vdots \\
    & \qw      & \qw      & \qw      & \ctrl{1}  & \qw    \\
    & \qw      & \qw      & \qw      & \targ     & \qw 
}
\end{equation}

The Hadamard test circuit depicted in Eq. \eqref{eq:htest_circ} requires the simulation
of controlled Pauli gadgets, however. Naively, this can be achieved by controlling
each single- and two-qubit gate in $V_1$, $V_2$ and $R_z(\theta_k)$ individually.
However, this would result in deep quantum circuits, given that the CNOT ladder transforms
into a CCNOT ladder. To decompose each CCNOT gate into single- and two-qubit gates
requires at least 6 CNOT gates, as well as additional Hadamard and T gates~\cite{shende2008cnot}.

By realizing that the Pauli gadget circuits in Eq. \eqref{eq:upg_circuit}
have a 'compute($V$)-action($A$)-uncompute($V^{\dagger}$)' structure,
the controlled evolution of $U_{\mathrm{Pg}}$ can be achieved
at much lower costs compared to naively controlling every gate
of $U_{\mathrm{Pg}}$:

Consider unitary matrices $U$, which can be decomposed as

\begin{equation}
U = V A V^\dagger \ ,
\end{equation}
where $V$ and $A$ are themselves unitary.
For such unitary matrices, the controlled operation $cU$ can be implemented by only
controlling the 'action', i.e. the unitary matrix $A$,
as indicated by Eq. \eqref{eq:conj_box_decomp}.

\begin{equation}\label{eq:conj_box_decomp}
cU =
\begin{pmatrix}
1 & 0 \\ 0 & U
\end{pmatrix}
= \begin{pmatrix}
1 & 0 \\ 0 & V A V^\dagger
\end{pmatrix}
= (1 \otimes V) \cdot cA \cdot (1 \otimes V^\dagger) \ .
\end{equation}

\begin{figure}[t]
\centering
\includegraphics[scale=0.65]{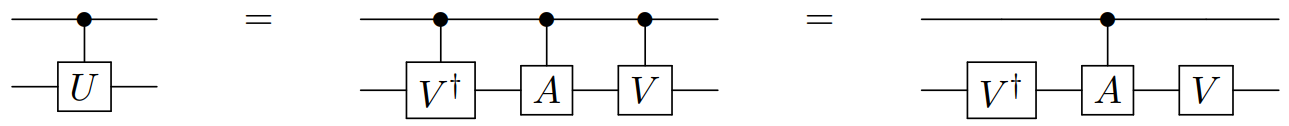}
\begin{comment}
\[
\Qcircuit @C=1.2em @R=1.0em {
    & \ctrl{1} & \qw \\
    & \gate{U} & \qw
}
\qquad = \qquad
\Qcircuit @C=1.2em @R=1.0em {
    & \ctrl{1} & \ctrl{1} & \ctrl{1} & \qw \\
    & \gate{V^\dagger} & \gate{A} & \gate{V} & \qw
}
\qquad = \qquad
\Qcircuit @C=1.2em @R=1.0em {
    & \qw & \ctrl{1} & \qw & \qw \\
    & \gate{V^\dagger} & \gate{A} & \gate{V} & \qw
}
\]
\end{comment}
\caption{Equivalent implementations of $cU$ for
$U=VAV^\dagger$.}
\label{fig:conj_box_decomp}
\end{figure}

A quantum circuit visualizing this identity is shown in Fig. \ref{fig:conj_box_decomp}.

Exploiting this identity can drastically reduce the circuit
depth, especially if the number of single- and two-qubit
gates required to implement $V$ compared to $A$ is large.
Looking at the quantum circuit decomposition for
$U_{\mathrm{Pg}}$ in Eq. \eqref{eq:upg_circuit},
we can identify the $U=VAV^\dagger$ structure with
$V= V_2^\dagger V_1^\dagger$, as well as $A=R_z(\theta_k)$.

This then enables the implementation of the circuit
of $cU_{\mathrm{Pg}}$ as

\begin{equation}\label{eq:cU_pg_circ}
cU_{\mathrm{Pg}} =
\Qcircuit @C=1.2em @R=1.0em {
   & \qw                        & \qw                        & \ctrl{2}           & \qw                & \qw                & \qw \\
   & \multigate{1}{V_1^\dagger} & \multigate{1}{V_2^\dagger} & \qw                & \multigate{1}{V_2} & \multigate{1}{V_1} & \qw \\
   & \ghost{V_1^\dagger}        & \ghost{V_2^\dagger}        & \gate{R_z(\theta)} & \ghost{V_2}        & \ghost{V_1}        & \qw
}
\end{equation}
at a greatly reduced circuit depth,
since only a single $R_z$ gate needs to be controlled
to implement $cU_{\mathrm{Pg}}$.

Further improvements in Trotterized time evolution circuit depths
can be achieved by considering the following two strategies,
which can also be combined:
(i) reordering the Trotter series and exploiting alternative synthesis strategies
of the CNOT ladder to maximize the number of CNOT cancellations between neighboring
Pauli gadgets.
(ii) Simultaneous diagonalization of sets of mutually commuting Pauli strings by Clifford circuits.
For an overview of circuit synthesis strategies following strategy (i) and (ii),
see e.g. Ref.~\cite{vuorenkoski2026shortest} as well as select examples
for strategy (i)~\cite{li2022paulihedral, de2024faster} and (ii)~\cite{van2020circuit, cowtan2020generic, mukhopadhyay2023synthesizing}.
For the quantum circuits that we compiled as detailed in the following
section \ref{sec:software_impl}, we mainly made use of
strategy (ii).

To employ strategy (ii), the Pauli strings $P_k$
in the qubit Hamiltonian (see Eq. \eqref{eq:H_qubit})
must be partitioned into sets of mutually commuting terms.
Determining the partitioning with the smallest number of sets corresponds to the
minimum clique cover problem, which is known to be NP-hard.
However, a plethora of heuristic algorithms with polynomial time complexity enable
efficient partitioning into a small number of cliques~\cite{verteletskyi2020measurement, reggio2024fast}.

\subsection{Software implementation and explicit gate counts}\label{sec:software_impl}

Our implementation of shallow controlled time-evolution circuits employs
the PennyLane~\cite{bergholm2018pennylane} and pytket~\cite{sivarajah2021t} libraries.
The PennyLane library is employed for preparation
of the (tapered) qubit Hamiltonians
and to sort its Pauli strings into mutually commuting sets by either a
qubit-wise commuting (QWC)~\cite{verteletskyi2020measurement},
or fully commuting (FCG) strategy~\cite{yen2020measuring}.
The pytket library is used to synthesize circuits, where Pauli gadgets
are either synthesized separately (SEP) using {\it PauliExpBox} objects,
or as mutually commuting sets, obtained by the QWC or FCG strategy,
where we employed {\it PauliExpCommutingSetBox} objects~\cite{cowtan2019phase, cowtan2020generic}.
As a final optimisation pass, the {\it SynthesiseTK} pass is utilised to compile
the circuits to generic single ({\it TK1})- and two ({\it TK2})-qubit gates,
where ${\it TK1}(\alpha, \beta, \gamma) = R_z(\alpha) R_x(\beta) R_z(\gamma)$,
and ${\it TK2}(\alpha, \beta, \gamma) = e^{-\frac{1}{2} i \pi \alpha (X \otimes X)} e^{-\frac{1}{2} i \pi \beta (Y \otimes Y)} e^{-\frac{1}{2} i \pi \gamma (Z \otimes Z)}$.

We do not employ any additional optimisation passes, e.g.
the {\it FullPeepholeOptimise} pass, which could decrease the circuit
depth by employing generic 2~\cite{vidal2004universal, blaauboer2008analytical}-
and 3~\cite{shende2005synthesis}-qubit unitary contractions.
This would have enabled the compilation of constant depth circuits as in Ref.~\cite{tranter2025high}
for the tapered two qubit Hamiltonian, for which we carried out
calculations on quantum hardware.
Since the contraction to constant depth circuits would have made simulations trivial,
and is no longer possible for larger than two qubit Hamiltonians, it was not employed here.

Employing the FCG strategy circuit compilation, we can compile quantum circuits
that implement a single Trotter time evolution step with 38 total (14 $TK2$ gates)
$TK1$+$TK2$ gates for the tapered two-qubit Hamiltonian that we employed for the
quantum hardware demonstration in section~\ref{quantinuum_results}.

To demonstrate the significant cost reduction of compiling circuits by a strategy
that exploits grouping of Pauli strings into mutually commuting sets,
we compare the number of quantum gates in circuits of a single controlled Trotterized time evolution step,
which employ the SEP, QWC, or FCG Pauli gadgetization in Table~\ref{tab:gate_cts_trotter_step_growth}.
Qubit tapering of the Hamiltonians was not utilized here, such that
the number of qubits in these circuits is $2n + 1$.
The number of active space electrons matches the number of active space spatial orbitals.

\begin{table}[h]
    \centering
    \begin{tabular}{c|c|c|c}\hline\hline
     Active space orbitals & FCG & QWC & SEP \\ \hline
     2 & 166 (73) & 177 (83) & 281 (136) \\
     4 & 2646 (1326) & 3370 (1988) & 6394 (3360) \\
     6 & 14253 (7724) & 21353 (14310) & 42979 (23496) \\
     8 & 48150 (27243)  & 81847 (58843) & 167719 (94592) \\
     10 & 122940 (71628) & 234731 (176254) & 487327 (280136) \\
     12 & 270263 (162822) & 555857 (432099) & - \\
     14 & 523840 (323005) & 1161722 (927284) & - \\
    \hline\hline
    \end{tabular}
    \caption{Total number of $TK1+TK2$ gates
    (only two-qubit $TK2$ gates) for a single Trotter step
    for an increasing number of orbitals in the active space.
    Comparison of circuit synthesis, where we make use of grouping
    Pauli strings into commuting sets (fully commuting grouping (FCG)
    and qubitwise commuting grouping (QWC)),
    and where Pauli string exponentials are compiled to the
    quantum circuit without first grouping them into commuting sets (SEP).}
    \label{tab:gate_cts_trotter_step_growth}
\end{table}

\subsection{A priori estimation of Trotter error}

As mentioned in section \ref{sec:trotter_scheme_det},
the ground state energy error due to Trotterization
can safely be estimated by considering upper bounds
of the form~\cite{childs2021theory}
$\lVert U_H^{(t)} - U_{\mathrm{Trot}}^{(t)} \rVert \leq \alpha(\lambda, t) t$,
where the polynomial in $t$ and the one-norm $\lambda$ depends on the
order of the employed product formula.

However, these bounds are known to be loose by multiple orders
of magnitude~\cite{reiher2017elucidating, mehendale2025estimating}, and would subsequently
lead to significantly deeper quantum circuits due to the use of larger Trotter step numbers
than required to reach the target energy error.
Selecting unnecessarily large Trotter step numbers is out of the question
for Hamiltonian simulation on the noisy quantum hardware that is available to us,
since too large Trotter step numbers lead to quantum circuit depths,
where errors due to low gate fidelities accumulate uncontrollably.
This would prohibit the accurate sampling of the autocorrelation function
that we require for QPD.

\begin{figure}[H]
    \centering
    \includegraphics[width=0.8\linewidth]{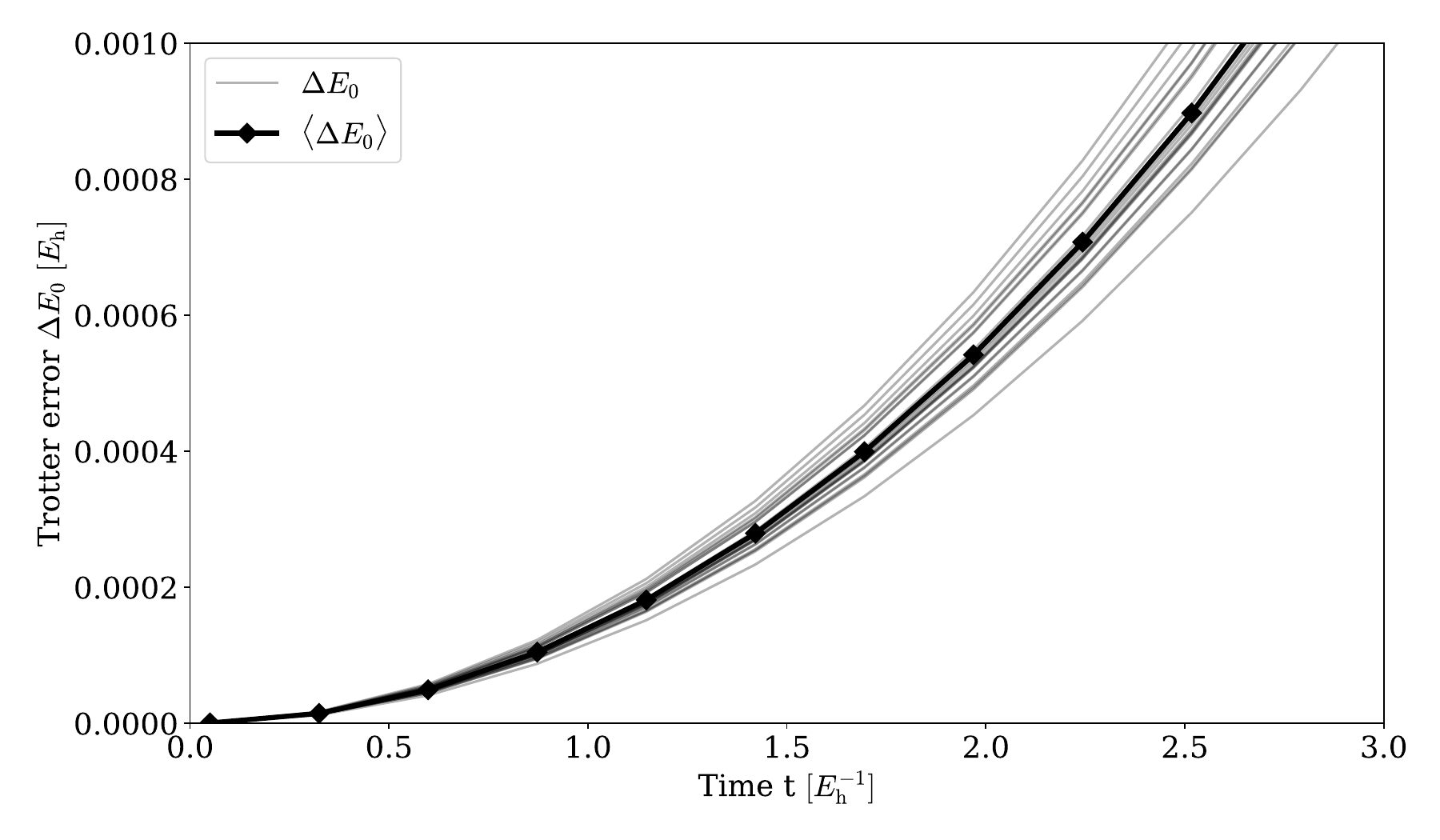}
    \caption{Ground state energy Trotter error as a function of the time evolution length $t$
    for 23 different bare CASCI (2o, 2e) Hamiltonians obtained from the bare structures of the ARG-TRP amino acid pairs.}
    \label{fig:arg_trp_trotter_error_growth}
\end{figure}

Consequently, we performed numerical calculations of the ground state energy
Trotter error on the set of bare ARG-TRP structures that were considered in section \ref{sec:sparse_sampling_qpd}.
The results are depicted in
Fig. \ref{fig:arg_trp_trotter_error_growth}.

From these results, we determined that a time evolution of $t=2.5 \ E_{\mathrm{h}}^{-1}$ per Trotter step
would be suitable for our quantum hardware calculations. 

\section{Estimation of ground state overlaps}
\label{sec:overlaps}

Quantum algorithms such as QPE and QPD benefit from large overlaps of the initial state with the ground state wave function.
We consider weakly correlated systems in this article, where the HF overlap is expected to be large.
In other words, even in the complete basis set (CBS) limit, the HF wave function $\ket{\psi_{\mathrm{HF}}}$ should have
significant, i.e. non-vanishing overlap with the exact (FCI) ground state wave function $\ket{\psi_0}$.

Assuming half-filling of the active space orbitals, FCI calculations are restricted to systems
with $\approx$ 20 spatial orbitals.
To estimate the overlap $\lvert \braket{\psi_0}{\psi_{\mathrm{HF}}} \rvert$ for larger systems,
we performed density matrix renormalization group (DMRG) calculation with the block2 program~\cite{zhai2023block2}.

The sweep algorithm described in Ref.~\cite{lee2021externally} and implemented in block2
was employed to obtain the HF coefficient in the CI expansion of the matrix product state (MPS) wave function.

\begin{figure}[H]
    \centering
    \includegraphics[width=\linewidth]{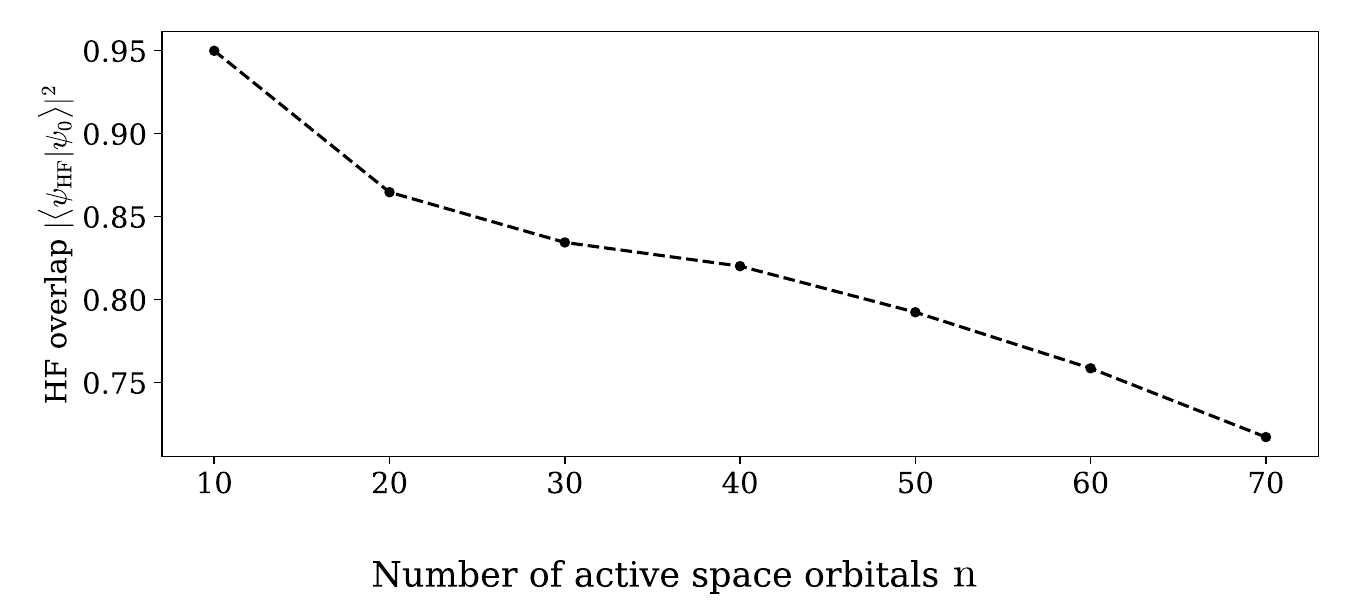}
    \caption{Squared overlap of the HF determinant with the ground state wave function
    as a function of the size of the active space.
    A DMRG wave function with a bond dimension of $\chi=1000$ serves
    as the reference $\ket{\psi_0}$.}
    \label{fig:hf_ovlp_embedded_hams}
\end{figure}

The HF overlap with an estimate of the ground state wave function for embedded QM/MM CASCI Hamiltonians of the ARG-TRP amino acid pair structure
is plotted as a function of
the active space size in Fig. \ref{fig:hf_ovlp_embedded_hams}.
The same molecular structure
for which we employed a quantum
computation with a minimal active space
was utilised.

Even for large active spaces with as many as 70 spatial orbitals
(mapping to 140 qubits after Jordan-Wigner transformation), substantial overlap
of the HF wave function and the ground state wave function is observed
($\lvert \braket{\psi_0}{\psi_{\mathrm{HF}}} \rvert^2 \approx 0.72$).
Consequently, the HF wave function serves as a suitable guiding state
for weakly correlated systems even at large active space sizes.
In the event that active spaces are so large
that the HF state is no longer a suitable guiding state,
even for a weakly correlated system, multi-determinant or MPS
guiding state preparation strategies can be adopted\,\cite{fomichev2024initial,erakovic2025high}.

\section{Mirror and randomized benchmarking of the Quantinuum
Helios noise model against the device}

Quantinuum's Helios processor is accompanied by an emulator
carrying a noise model of the device~\cite{ransford2026}.
The published characterization of Helios reports component
errors from Clifford randomized benchmarking and a system-level
random-circuit fidelity inferred from mirror
benchmarking~\cite{ransford2026,decross2025,mayer2023}.
We reproduce both at four qubits and run the identical circuits on the emulator and on the device.
Two sets of benchmark circuits were used: mirror
circuits and randomized benchmarking circuits.

In mirror circuits,
a \emph{layer} on $n$ qubits consists of a Haar-random
single-qubit gate on every qubit, followed by
$\mathrm{ZZPhase}(\pi/2)$.
Each layer carries $n$ single-qubit gates and $n/2$ entanglers.
A circuit of layer depth $l$ executes $l/2$ layers
$L_1, \ldots, L_{l/2}$ and then their inverses in reverse order,
$L_{l/2}^{-1}, \ldots, L_1^{-1}$.
Each circuit is initialized in a random computational basis
state.

For randomized-benchmarking circuits,
a sequence of length $m$ applies $m$ Cliffords drawn uniformly
from the $k$-qubit Clifford group, followed by the single
Clifford inverting their product, computed by composing
rather than by inverting circuits:
\begin{equation}
  C_{\mathrm{inv}} = \left( C_m \cdots C_2 C_1 \right)^{-1}.
  \label{eq:rbinv}
\end{equation}
In the absence of noise, 
the final state is $\lvert 0 \ldots 0 \rangle$.
The benchmark measures the fraction of shots returning that bitstring.
Circuits are constructed in the Helios native gate set (PhasedX, Rz, ZZPhase) without any
optimisation pass.

The four-qubit mirror
circuits had depths $l \in \{2, 4, 8, 16\}$ with
five circuit instances each.
One-qubit randomized benchmarking used length
$m \in \{1, 8, 64\}$,
and two-qubit used $m \in \{1, 4, 16\}$;
three sequences in each case.
Thirty-eight circuits  per target, 
were submitted in ten jobs of at
most four programs.
The emulator pass ran at
$100$ shots per circuit and the device pass at
$50$ shots. Moreover, to further validate the scaling analysis presented in Figg.\,\ref{fig:noise_scaling_small_active_space}-\ref{fig:noise_scaling_large_active_space}, we executed the same set of circuits also with the noise model built through the PECOS library\,\cite{pecos} based on the hardware specifics publicly available at\,\cite{heliosparams}. 
The results are summarized in Table~\ref{tab:comparison}. 

\begin{table}[h!]
\caption{Probability of the ideal outcome per cell for the offline noise model
(\texttt{PECOS}), the emulator (\texttt{Helios-1E}) and the device
(\texttt{Helios-1}), at 100, 100 and 50 shots per circuit.  Uncertainties are standard errors across circuit instances. The last column is
the emulator--device difference in units of its combined standard error.}
\label{tab:comparison}
\begin{ruledtabular}
\begin{tabular}{lcccc}
cell  & \texttt{PECOS} & \texttt{Helios-1E} & \texttt{Helios-1} & $n_\sigma$ \\
\colrule
mirror $l=2$   & $0.998(3)$  & $0.992(6)$  & $0.988(8)$  & $0.40$ \\
mirror $l=4$   & $0.988(8)$  & $0.996(2)$  & $0.988(12)$ & $0.65$ \\
mirror $l=8$   & $0.978(9)$  & $0.980(6)$  & $0.992(5)$  & $1.50$ \\
mirror $l=16$  & $0.962(9)$  & $0.960(7)$  & $0.944(15)$ & $0.98$ \\
\colrule
2Q RB $m=1$    & $1.000(0)$  & $0.997(3)$  & $1.000(0)$  & $1.00$ \\
2Q RB $m=4$    & $0.987(9)$  & $0.987(3)$  & $0.993(7)$  & $0.89$ \\
2Q RB $m=16$   & $0.967(14)$ & $0.970(12)$ & $0.953(24)$ & $0.62$ \\
\colrule
1Q RB, all $m$ & $1.000(0)$  & $1.000(0)$  & $1.000(0)$  & --- \\
\end{tabular}
\end{ruledtabular}
\end{table}

\section{Execution on Quantinuum Helios}
\label{app:helios}

Circuits were delivered as serialised \texttt{pytket} dictionaries and executed with  a Python driver built on
\texttt{pytket}, \texttt{guppylang}/\texttt{hugr} and
\texttt{qnexus}.
Each circuit was compiled locally to the native gate set (\textsc{PhasedX}, $R_z$,
\textsc{ZZPhase}).
Optimisation level~1 was used for the
three-qubit QMMM sets, since level~2 would compress every QMMM circuit to constant-depth and removed the depth
dependence on $t$.

Compiled packages were uploaded once and were submitted as a single multi-program batch
job.
Before execution, the
same uploads ran on the free syntax checker \texttt{Helios-1SC}.
Per program cost obtained from the syntax checker was doubled
te set the execution cap.
Job size was set by the backend.  
In August 2026, the 256 circuits of the QMMM sets were executed 
as 16 jobs of 16 programs each
on the Helios-1E emulator at 100 shots per circuit, and on the Helios-1 processor at 10 shots per circuit.

Submission and collection were separate asynchronous steps.  Job ids, program ids,
statuses, costs and count histograms were written to an atomic,
resumable JSON worksheet before polling; a fetch loop polled per-item
status until all items of a job were terminal, then downloaded the
counts.  Failed items were resubmitted from the
stored program ids without recompilation; items completed on Nexus
but not downloaded were fetched in a repair pass.

\section{Hankel Structured Newton Like Descent (HSNLD)}
\label{app:hsnld}

In this section, we provide the technical details of the HSNLD algorithm to complement the discussion in the main text. For further details, we refer the reader to Ref.\,\cite{Cai2025}. Particularly, here we discuss the update rule of the gradient descent algorithm, including its preconditioner and projection, which enable faster convergence even when the underlying Hankel matrix is ill-conditioned. We start describing the initialization strategy adopted by the authors.

\subsubsection*{Initialization}

The HSNLD algorithm operates on a reweighted form of the signal vector $\textbf{y} = \textbf{W}^{-1}\textbf{x}$ where $\textbf{W}=\text{diag}(\zeta_j^{-1/2})$ with $\zeta_j$ being the number of entries on the j-th skew diagonal (see Sec.\,\ref{sec:cs}). Analogously we can define the reweighted Hankel map as $\mathcal{G}=\Hank\textbf{W}$ such that $\mathcal{G}(\textbf{y})=\Hank(\textbf{x})$. Having in place this reweighted notation we can write the initial guess of the HSNLD optimization as

\begin{equation}
U_0\Sigma_0V_0^\ast=\mathrm{SVD}_r\Bigl[\frac{1}{p_{\Om}}\,\mathcal G\,\PiOm\tilde{\mathbf y}\Bigr],\qquad
\bigl(L^{(0)},R^{(0)}\bigr)=\mathcal P_{\mathrm{inc}}\bigl(U_0\Sigma_0^{1/2},\,V_0\Sigma_0^{1/2}\bigr),
\label{eq:init}
\end{equation}
where $\mathrm{SVD}_r$ is the singular value decomposition truncated to the $r$ largest singular values.

The projection $\mathcal P_{\mathrm{inc}}$ keeps the factors incoherent where incoherence measures how much the information is spread through the measurement set in a compressed sensing setting. Let $\Hank(\mathbf x)=U\Sigma V^\ast$ be the singular value decomposition of $\Hank(\mathbf x)$, with $U\in\Cc^{N_1\times r}$, $V\in\Cc^{N_2\times r}$ and singular values $\sigma_1(\Hank(\mathbf x))\ge\dots\ge\sigma_r(\Hank(\mathbf x))$. The matrix $\Hank(\mathbf x)$ is $\mu_{\Hank}$-incoherent if
\begin{equation}
\norm{U}_{2,\infty}\le\sqrt{\frac{\mu_{\Hank}\varrho r}{N}},\qquad
\norm{V}_{2,\infty}\le\sqrt{\frac{\mu_{\Hank}\varrho r}{N}},\qquad
\varrho=\max\Bigl\{\frac{N}{N_1},\frac{N}{N_2}\Bigr\},
\label{eq:incoherence}
\end{equation}
where $\norm{\cdot}_{2,\infty}$ is the largest row norm. A value of $\mu_{\Hank}$ close to one means that $\Hank(\mathbf x)$ is spread evenly over its rows and columns, so that no entry of $\mathbf x$ carries a large share of the information. Conversely, high values of $\mu_{\Hank}$ denote matrices where the information is very localized over few data points.

Using explicitly the reweighted notation we can write the cost function of the HSNLD algorithm as 

\begin{equation}
\ell(L,R)=\frac{1}{2p}\bigl\langle\PiOm(\mathcal G^\ast(LR^\ast)-\mathbf f),\,\mathcal G^\ast(LR^\ast)-\mathbf f\bigr\rangle
+\frac12\norm{(I-\mathcal G\mathcal G^\ast)(LR^\ast)}_F^2 .
\label{eq:objective}
\end{equation}

Where the rank constraint is enforced by factoring the Hankel matrix as $LR^\ast$ with $L\in\Cc^{N_1\times r}$, $R\in\Cc^{N_2\times r}$, and
the Hankel structure by a penalty.

\subsubsection*{Gradient descent update rule}

HSNLD minimizes
Eq.~\ref{eq:objective} by preconditioned gradient steps,
\begin{equation}
\begin{pmatrix}L_{k+1}\\ R_{k+1}\end{pmatrix}
=\mathcal P_{\mathcal C}
\begin{pmatrix}
L_k-\eta\,\nabla_L\ell(L_k,R_k)\,(R_k^\ast R_k)^{-1}\\[2pt]
R_k-\eta\,\nabla_R\ell(L_k,R_k)\,(L_k^\ast L_k)^{-1}
\end{pmatrix},
\label{eq:hsnld-step}
\end{equation}
with step size $\eta\in(0,1]$. The factors $(R_k^\ast R_k)^{-1}$ and
$(L_k^\ast L_k)^{-1}$ are the Newton-like preconditioners introduced in Ref.\,\cite{tong2021accelerating} which make the
convergence rate independent of the condition number $\kappa$ of
$\Hank(\mathbf x)$, whereas plain gradient descent on the same objective
requires a number of iterations growing linearly with $\kappa$\,\cite{cai2023structured}.

Ref.\,\cite{Cai2025} proves that the optimization starting from Eq.~\eqref{eq:init} with the conditioned update of \,\eqref{eq:hsnld-step} recovers the underlying $\Hank(\mathbf x)$ if the number of samples satisfies
\begin{equation}
|\Om^{\pm}|\;\gtrsim\;\varrho^2\mu_{\Hank}^2 r^2\kappa^4\log N.
\label{eq:sample-complexity}
\end{equation}

Finally, we now connect Eq.~\eqref{eq:sample-complexity} to the number of executed points, Eq.~\eqref{eq:keep}. For the rank $r\approx1$ used in this work, $\kappa\approx1$, and for a single node on the unit circle $\mu_{\Hank}\approx1$. For the square shape $N_1=N_2=N_s+1$ of Sec.~\ref{sec:cs}, $\varrho=(2N_s+1)/(N_s+1)<2$. With $|\Om^{\pm}|=2|\Om|+1$ and $N=2N_s+1$, Eq.~\eqref{eq:sample-complexity} becomes
\begin{equation}
|\Om|\;\ge\;\frac{c_{\Om}\varrho^2}{2}\,\ln N_s,
\label{eq:keep-from-thm}
\end{equation}
where the base of the logarithm has been absorbed into $c_{\Om}$. Eq.~\eqref{eq:keep} has the same form, with $C_{\mathrm{cs}}$ in place of $c_{\Om}\varrho^2/2$. In our work the constant $C_{\mathrm{cs}}$ is fixed empirically (Sec.~\ref{sec:methods}).

\section{The Arginine-Tryptophan system: rationale of the choice for this system and the generation of the structures. }
\label{app:molecular}

\subsection{Amino acids pairs as use-case systems.}
\begin{figure}[h!]
    \centering
    \includegraphics[width=0.75\linewidth]{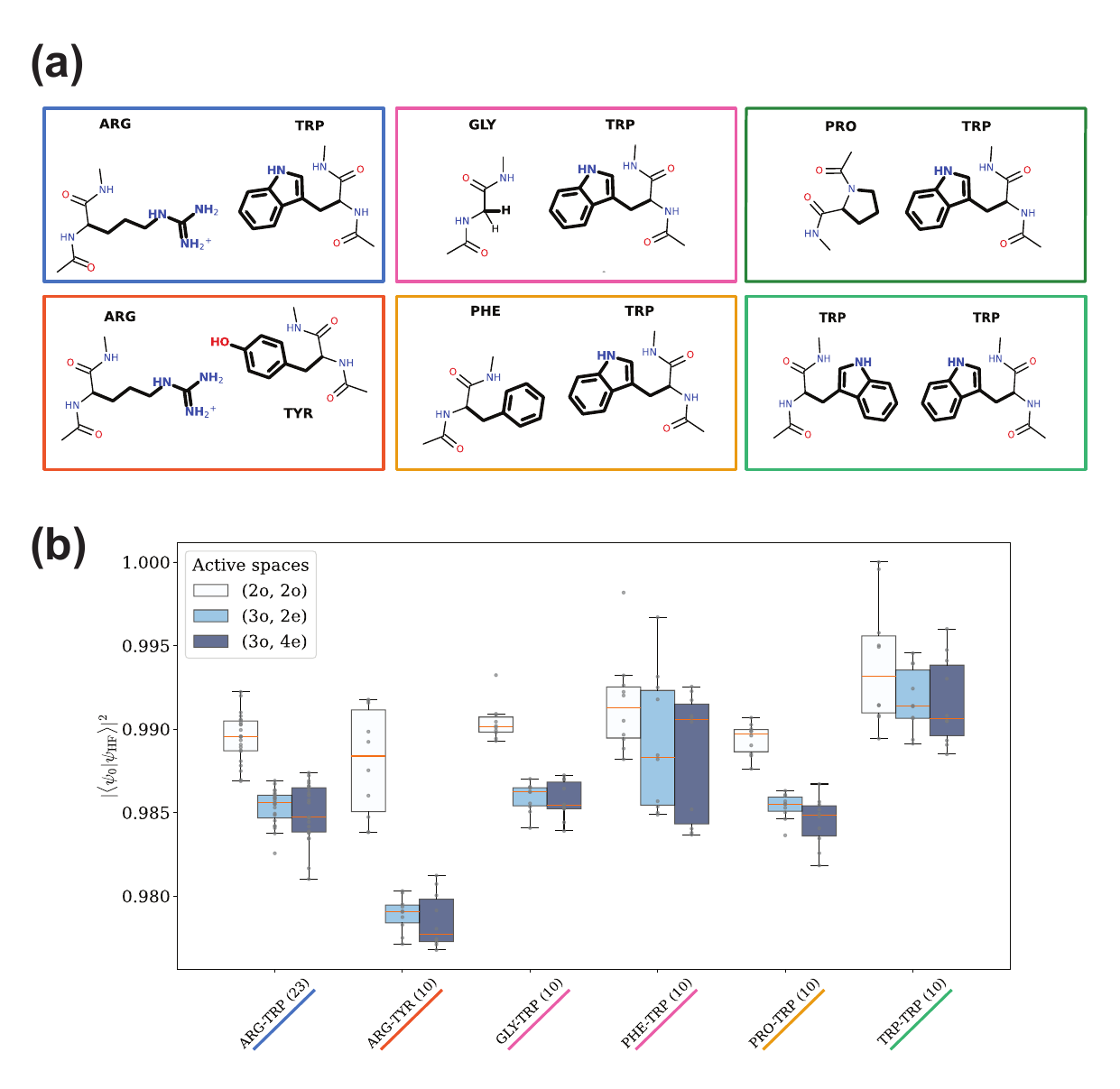}
    \caption{Amino acid pairs studied and multireference character of their interactions.
(a) Chemical structures of the six amino acid residue pairs considered, with each pair represented also by a color, and with side chains highlighted in bold.
(b) Box plots of the overlaps of the Hartree--Fock wave function $\ket{\Psi_{\mathrm{HF}}}$ with the CASCI ground-state wave function $\ket{\Psi_0}$ for CASCI Hamiltonians of varying active space sizes, computed on representative molecular structures. The number of structures evaluated for each amino acid pair is indicated in brackets. Individual data points are shown as gray circles.}

    \label{fig:pmfs}
\end{figure}

Proteins are polymers of amino acids, consisting of a common polypeptide backbone (the main chain of the polymer) bearing a set of branched groups called side chains. Side chains are of 20 canonical types (see Figure~\ref{fig:pmfs}a for a visualization of the chemical structures of six amino acids, with their side chains highlighted in bold; further, the backbone has been modified (`capped') so as to represent the situation within a protein). Understanding the behavior of these biomolecules, and thereby characterizing their chemical and structural properties, requires a thorough understanding of pairwise amino acid interactions. 

The chemistry of the side chain governs the nature of these interactions: including hydrogen bonds for polar side chains, electrostatics for charged side chains, and through $\pi$ orbitals for aromatic groups. This latter class of interactions is of major interest, as it is thought to drive certain biomolecular phenomena such as the phase separation of biomolecular condensates~\cite{vernon_pi-pi_2018, das_comparative_2020, bremer_deciphering_2022}; nevertheless even in state-of-the-art molecular models these are generally described with empirical force fields that must be used when simulating biomolecular complexes of critical size~\cite{ivanovic2025all}. This motivates the examination of these systems at higher accuracy, without approximations on their quantum description.

In this work, we considered six amino acid pairs: arginine-tryptophan (ARG-TRP), glycine-tryptophan (GLY-TRP), proline-tryptophan (PRO-TRP), arginine-tyrosine (ARG-TYR), phenylalanine-tryptophan (PHE-TRP), and tryptophan-tryptophan (TRP-TRP) (Fig.~\ref{fig:pmfs}a). In each of these pairs, at least one of the two amino acid residues carries an aromatic side chain. These six pairs were selected from the 210 possible pairs arising from 20 amino acids because (1) they involve a $\pi$ system, and (2) their favorable interactions have been shown to play a major role in biomolecular self-assembly~\cite{wang2018molecular,bremer_deciphering_2022}.

\subsection{Sampling of the pair interaction with all atom molecular dynamics.}

Umbrella sampling molecular dynamics simulations were conducted using OpenMM 8.4~\cite{eastman_openmm_2024} patched with Plumed \cite{tribello_plumed_2014} at the all-atom level using the Amber ff19SB force field~\cite{tian_ff19sb_2020} in explicit water (OPC model~\cite{izadi_building_2014}) to sample the interactions of the investigated pairs. The systems consist of two amino acids capped to neutralize the charged backbone groups (capping groups were ACE and NME, and the structures were built with Ambtertools~\cite{case_ambertools_2023}). Each system was solvated in a 0.15~M NaCl solution, energy-minimized, and equilibrated (10~ns of NPT molecular dynamics). Eighteen replicas of the simulation, also called eighteen windows of the umbrella sampling, were run in parallel, with harmonic restraints on the side-chain center-of-mass--side-chain center-of-mass (scCOM--scCOM) distance, centered on evenly distributed values ranging from 0.3~nm to 1.8~nm.

\subsection{Extraction of interaction representative structures through clustering analysis.}\label{sec:struc_extraction}

The umbrella sampling molecular dynamics simulations serve as the basis for extracting representative structures for each amino acid pair considered. Their trajectories were first filtered to retain only `in-contact' frames, defined as structures for which at least one interatomic distance between the two amino acids is less than 5~\AA, and then clustered using the YACARE algorithm~\cite{descamps2025clustering}. This algorithm explores the density from frame to frame in a given observable space; in our case, the distances between inter-amino acid hydrogen-bond partners and $\pi$-system atoms.

The YACARE algorithm was used because it enables the identification of stable interaction modes consistently adopted by each pair. For each pair, the representative frames of each cluster were extracted and the solvent molecules removed, yielding a set of molecular structures used to benchmark our quantum computing algorithm. The number of representative frames for each pair is indicated in Fig.~\ref{fig:pmfs}b.
The cluster representatives of the ARG-TRP pair are shown in the reduced observable space of the pair’s trajectory, projected onto the first and second principal components, in Figure~\ref{fig:clustering} of the main text.

\subsection{Choosing one pair as use-case system: arginine-tryptophan.}

For our demonstration of QPD on quantum hardware, the selection
of one of the six initially considered systems was based on the following two criteria:
(i) the ground state overlap of the HF state should be comparatively
small. This generally leads to a larger correlation energy,
and, hence, places higher demands on the energy improvements
for QPD to reach the chemical accuracy target.
(ii) The selection of a consistent active orbital space
for different molecular structures obtained by the clustering analysis
should be possible.

We assessed the ground state overlap of the HF state
for these structures and for active space sizes,
which were initially considered for the QPD calculation on
the Quantinuum Helios quantum computer.
The results are visualized in Fig.~\ref{fig:pmfs}b.

\begin{figure}[h]
    \centering
    \includegraphics[width=0.85\linewidth]{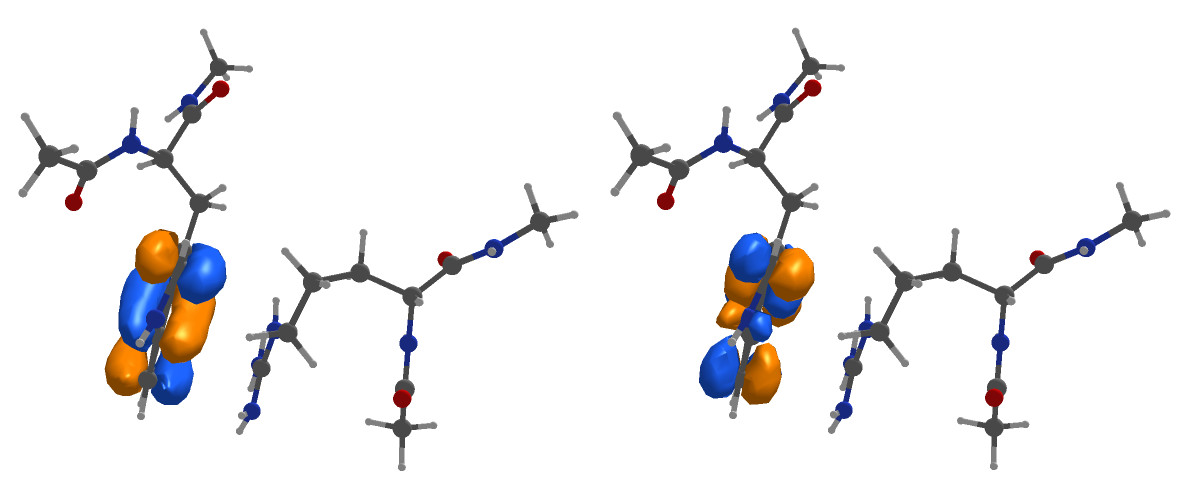}
    \caption{HOMO (left) and LUMO (right) orbitals of an example
    structure of an isolated ARG-TRP amino acid pair.}
    \label{fig:homo_lumo_arg_trp_orb_vis}
\end{figure}

Due to these criteria, we selected the ARG-TRP
pair system for this manuscript.
While the ARG-TYR system
displays slightly smaller overlaps of the HF guiding state
with the ground state, the MP2 natural orbitals
around the HOMO-LUMO level did not display a consistent
ordering among different molecular structures.
Consequently, additional effort would have been required
for consistent orbital selection, e.g. by employing
(semi-)automated orbital selection
schemes~\cite{sayfutyarova2017automated, stein2019autocas, tarocco2025aegiss}.
For simplicity, we therefore opted for the ARG-TRP system,
where the bonding and anti-bonding $\pi$ orbitals located
on the tryptophan side chain were consistently ordered
across various inspected structures.
See Fig.~\ref{fig:homo_lumo_arg_trp_orb_vis} for a visualization example
of the HOMO and LUMO orbitals of an example structure of the ARG-TRP pair.
Additionally, we highlight that arginine and tryptophan engage in a cation-pair interactions, which are major contributions to the physicochemical process of phase separation~\cite{nott2015phase,wang2018molecular}.

\subsection{QM/MM system considered for quantum hardware simulation}\label{sec:qm_mm_sys}

\begin{figure}[h]
    \centering
    \includegraphics[width=0.65\linewidth]{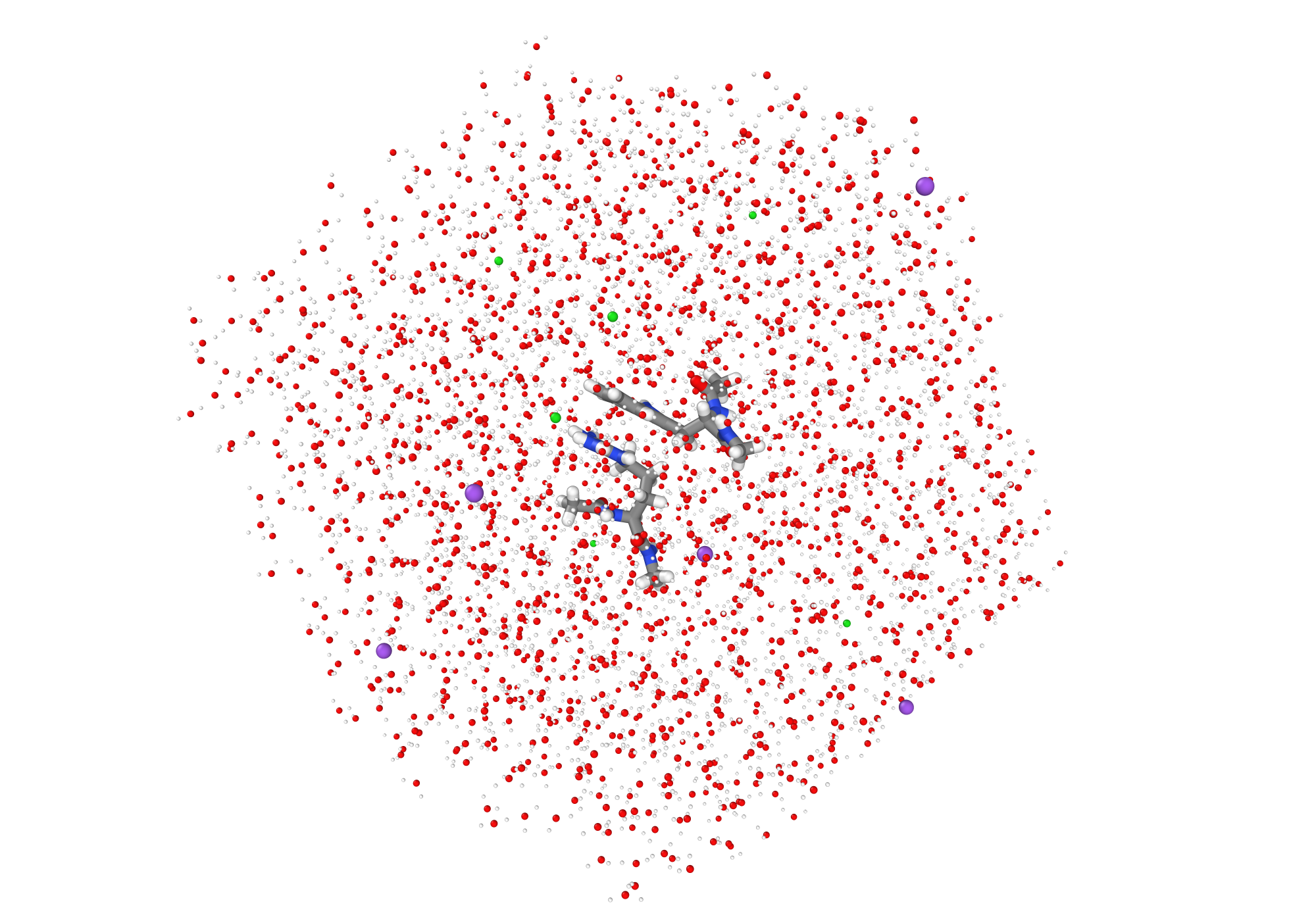}
    \caption{MD simulation snapshot of the ARG-TRP pair, surrounded
    by 1871 water molecules, as well as 6 chloride and 5 sodium ions.
    The total system thus carries a net charge of zero.
    }
    \label{fig:snapshot_embedded_sys}
\end{figure}

The molecular structure, which corresponds to a snapshot
of a randomly chosen QM/MM production run frame,
from which we obtained a QM/MM
embedded CASCI Hamiltonian,
as described in section~\ref{sec:methods},
and which was subsequently deployed for the QPD
demonstration on quantum hardware in section~\ref{quantinuum_results}, is depicted in Fig.~\ref{fig:snapshot_embedded_sys}.

For the QM/MM simulation, the use of the OPC water model!\cite{izadi_building_2014} determines
coordinates and point charges required for the embedded
Hamiltonian in Eq.~\eqref{eq:H_qmmm}.
The point charges for the chloride and sodium ions were
set to $-1$, and $+1$, respectively. 

\section{Cramer-Rao lower bounds for sparse QPD}
\label{app:crlb}

In this section we derive lower bounds on the error estimate for QPD based on the Cramer-Rao Lower Bound\,\cite{cramer1999mathematical, rao1992information} (CRLB). We closely follow the work of Ref.\,\cite{dutkiewicz2022heisenberg,kimura2026fisher}. It is important to notice that this lower bound on the mean squared error derives uniquely from the acquisition design of the data and so applies to all methods that share it. Analogously, it is completely agnostic of the post-processing which may affect the final scaling of the algorithm. A final important remark that we have to make before proceeding is that this type of error bounds formally apply only when $\hat{E}$ is an unbiased estimator of $E$. We will go back to this point at the end of our derivation. 

For an unbiased estimator $\hat E_0(\mathbf x)$ of the ground-state energy, the CRLB states
\begin{equation}
\label{crlb_def}
    \epsilon^2=\sum_{\mathbf x}\mathbb P(\mathbf x|\boldsymbol\varphi)\bigl(\hat E_0(\mathbf x)-E_0\bigr)^2\ \ge\ \bigl(I^{-1}\bigr)_{E_0E_0}\
\end{equation}
where $I$ is the Fisher information matrix (FIM) of the data, the data $\mathbf x$ depend on the parameters $\boldsymbol\varphi=(\mathbf E,\mathbf p)$, the energies $E_m$ and the populations $p_m=a_m^2$.
The matrix elements of $I$ are given according to the following expression\,\cite{dutkiewicz2022heisenberg}
\begin{equation}
    I_{ij}(\boldsymbol\varphi)=\sum_{\mathbf x}\mathbb P(\mathbf x|\boldsymbol\varphi)\,
    \partial_{\varphi_i}\ln\mathbb P(\mathbf x|\boldsymbol\varphi)\,
    \partial_{\varphi_j}\ln\mathbb P(\mathbf x|\boldsymbol\varphi).
\end{equation}

The idea behind this derivation is that the total FIM is given by the sum of the FIM related to the computation of a single sample of $C(t)$ since the FIM for independent measurements is additive. Following Ref.\,\cite{kimura2026fisher} we use $(I_{E_0E_0})^{-1}$ as a proxy for $(I^{-1})_{E_0E_0}$, i.e. we assume 
\begin{equation}
\label{km_assumption}
    (I_{E_0E_0})^{-1} \approx (I^{-1})_{E_0E_0}.
\end{equation}
The authors of Ref.\,\cite{kimura2026fisher} have shown that this holds when the minimal energy gap $\Delta$ in $C(t)$ is positive and $T_{\max} \gg \Delta^{-1}$.

Each shot of the Hadamard test at time $t_k$ returns $\sigma=\pm1$, and the real- and imaginary-part circuits (Sec.\,\ref{sec:hadamard}) are run independently. Writing $C(t_k)=X(t_k)+iY(t_k)=\sum_mp_m\bigl(\cos E_mt_k-i\sin E_mt_k\bigr)$, for an algorithm collecting $M_k$ shots of each circuit at times $t_k$ we have
\begin{equation}
\label{FMI_total}
    I_{ij}=\sum_kM_k\,I^{C}_{ij}(t_k),\qquad
    I^{C}_{ij}(t_k)=I^{X}_{ij}(t_k)+I^{Y}_{ij}(t_k),
\end{equation}
where $I^{X(Y)}(t_k)$ is the FIM of one shot of the real- (imaginary-) part circuit. For the real part, Eq.\,\eqref{eq:Hadamard-re} gives $\mathbb P(\sigma|\boldsymbol\varphi,t_k)=\frac12\bigl[1+\sigma X(t_k)\bigr]$, so that $\partial_i\mathbb P=\frac\sigma2\,\partial_iX$ and
\begin{equation}
\label{FMI_single_sample}
    I^{X}_{ij}(t_k)=\sum_{\sigma=\pm1}\frac{\partial_i\mathbb P(\sigma)\,\partial_j\mathbb P(\sigma)}{\mathbb P(\sigma)}
    =\frac{\partial_iX\,\partial_jX}{4}\Bigl[\frac{2}{1+X}+\frac{2}{1-X}\Bigr]
    =\frac{\partial_iX\,\partial_jX}{1-X(t_k)^2},
\end{equation}
and in the same way $I^{Y}_{ij}(t_k)=\partial_iY\,\partial_jY/\bigl(1-Y(t_k)^2\bigr)$. The derivatives are
\begin{align*}
  \frac{\partial X}{\partial E_j}&=-p_j\,t_k\sin E_jt_k, &
  \frac{\partial Y}{\partial E_j}&=-p_j\,t_k\cos E_jt_k,\\
  \frac{\partial X}{\partial p_j}&=\cos E_jt_k, &
  \frac{\partial Y}{\partial p_j}&=-\sin E_jt_k ,
\end{align*}

Substituting Eq.\,\eqref{FMI_single_sample} into Eq.\,\eqref{FMI_total} we get

\begin{equation}
\label{starting_FIM}
    I_{E_0E_0} = \sum_k M_k p^2_0 t^2_k F_0(t_k) = p^2_0 \sum_k M_k t^2_k [\frac{\sin^2 E_0t}{1-X(t)^2}+\frac{\cos^2 E_0t}{1-Y(t)^2}]
\end{equation}
This last equation already can be applied to bound (under the assumption of Eq.\,\eqref{km_assumption}) the error estimate $\epsilon$. We now proceed specializing the previous equation for the case of (sparse) QPD.

In particular, we know that for standard QPD we need to collect $N_s = \lceil W_s T_{max}/\pi\rceil$ samples using $M_s=F\sqrt{N_s\log N_s}$ shots (in the following we will drop for simplicity the dependence on $F$) with a timestep $\delta t = \frac{\pi}{W_s}$. Then, we obtain

\begin{equation}
  T_{max}=\frac{\pi N_s}{W_s},\qquad
  T_{runtime}=2M_s\,\frac{\pi}{W_s}\sum_{k=1}^{N_s}k=M_s\,\frac{\pi}{W_s}\,N_s(N_s+1).
  \label{eq:QPD_cost}
\end{equation}

Substituting into Eq.\,\eqref{starting_FIM} we obtain

\begin{equation}
\label{FIM_QPD}
    I_{\mathrm{QPD}}=M_s\,p_0^2\,\frac{\pi^2}{W_s^2}\sum_{k=1}^{N_s}k^2F_0(t_k) = M_s\,p_0^2\,\frac{\pi^2}{W_s^2}\,\bar{F}\sum_{k=1}^{N_s}k^2
\end{equation}
where $\bar{F}=\sum_{k=1}^{N_s}w_k\,F_0(t_k),\, w_k=\frac{k^2}{\sum_{l=1}^{N_s}l^2},\, \bar{F}\in\Bigl[1,\frac2{p_0}\Bigr]$. Please notice also that the index $k$ starts from 1 as we do not compute the first initial point as it is always $C(0)=1$. Using $\sum_{k=1}^{N}k^2=N(N+1)(2N+1)/6$ we finally get

\begin{equation}
      I_{\mathrm{QPD}}=M_s\,p_0^2\,\bar{F}\,\frac{\pi^2}{W_s^2}\,\frac{N_s(N_s+1)(2N_s+1)}{6}
\end{equation}
the previous equation leads to the bound

\begin{equation}
    \epsilon^2 \ge\ I_{QPD}^{-1} \ge\ \frac{3\,W_s^2}{\pi^2M_s\,p_0\,N_s(N_s+1)(2N_s+1)}.
\end{equation}
where the last inequality is obtained substituting $\bar{F}=\frac{2}{p_0}$.

To obtain an asymptotic scaling we consider the limit of $N_s\to\infty$ at fixed $W_s$. In order to proceed we adopt the following approximations, $N_s(N_s+1)(2N_s+1)/6\approx\tfrac13N_s^3[1+O(N_s^{-1})]$ and $N_s(N_s+1)\approx N_s^2[1+O(N_s^{-1})]$. In this limit we get

\begin{equation}
  \epsilon^2 \geq \frac{3\pi}{M_s\,p_0^2\,\bar{F}\,W_s\, T_{max}^3},\,\,
  T_{runtime}=\frac{M_s\,W_s\,T_{max}^2}{\pi}.
  \label{eq:asym_general}
\end{equation}
substituting $M_s= (W_s\Tmax/\pi\ln(W_s\Tmax/\pi))^{1/2} = (W_s\Tmax/\pi)^{1/2}L^{1/2}$ reads

\begin{equation}
  \label{eq:asym_Tmax}
  \epsilon^2 \simeq\frac{3\pi^{3/2}}{\,p_0^2\,\bar{F}}\;W_s^{-3/2}\,T_{max}^{-7/2}\,L^{-1/2} = \frac{3}{p_0^2\,\bar{F}}\Bigl(\frac{W_s}{\pi}\Bigr)^{3/5}L^{1/5}\;T_{runtime}^{-7/5}
\end{equation}
with
\begin{equation}
    T_{runtime}\simeq\frac{1}{\pi^{3/2}}\;W_s^{3/2}\,T_{max}^{5/2}\,L^{1/2}.
  \label{eq:asym_Truncost}  
\end{equation}

Finally, up to logarithmic factors we get

\begin{equation}
    \epsilon\propto\Tmax^{-7/4}\propto T_{runtime}^{-7/10}.
\end{equation}

In the next section we modify this derivation that applies for the QPD estimator, to account for sparse sampling.

\subsection{Accounting for sparse sampling}

\begin{figure}[ht]
\centering
\includegraphics[scale=0.65]{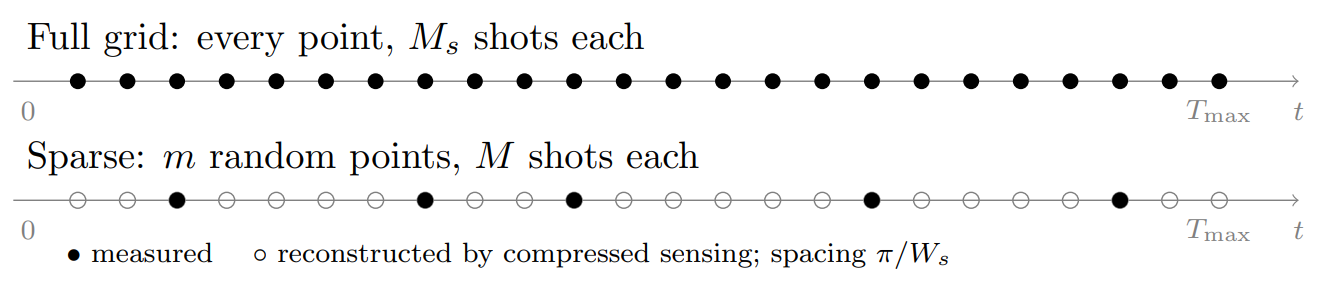}
\begin{comment}
\begin{tikzpicture}[x=0.5cm,y=0.5cm]
  \foreach \row/\lab in {0/{Full grid: every point, $M_s$ shots each},-2.4/{Sparse: $m$ random points, $M$ shots each}}{
    \draw[->,gray] (-0.3,\row) -- (25.6,\row);
    \node[anchor=west,font=\small] at (-0.3,\row+0.85) {\lab};
    \node[font=\scriptsize,gray] at (0,\row-0.6) {$0$};
    \node[font=\scriptsize,gray] at (24,\row-0.6) {$\Tmax$};
    \node[font=\scriptsize,gray] at (25.6,\row-0.6) {$t$};
  }
  \foreach \k in {1,...,24}{ \fill[black] (\k,0) circle (0.16); }
  \foreach \k in {1,...,24}{ \draw[gray] (\k,-2.4) circle (0.16); }
  \foreach \k in {3,8,11,17,22}{ \fill[black] (\k,-2.4) circle (0.16); }
  \node[anchor=west,font=\scriptsize] at (0.5,-3.5) {$\bullet$ measured \quad $\circ$ reconstructed by compressed sensing; spacing $\pi/W_s$};
\end{tikzpicture}
\end{comment}
\caption{The two acquisition designs on the same grid $t_k=\pi k/W_s$.}
\label{fig:designs}
\end{figure}

In this section we drop the requirement of measuring every point on the QPD sampling grid and adopt the sparse sampling allowed by HSNLD as proposed in this work measuring $m\ll N_s$ grid points, chosen at random\,\ref{fig:designs}. Moreover we assume that the number of shots per sample does not depend on $N_s$ as in the previous section, therefore we write $M$ instead of $M_s$, Fig.\,\ref{fig:designs}.

To account for the previous statement we need compute the total FIM of Eq.\,\eqref{FIM_QPD} substituting to $\sum_k k^{2}$ the average of the sum of the drawn samples. From Ref.\,\cite{rice2007mathematical}(Ch. 7.3) we have

\begin{equation}
    \mathbb{E}[\sum_{j=1}^m t_{k_j}^2] = \frac{m}{N_s} \frac{\pi^2}{W_s^2} \sum_{k=1}^{N_s} k^2 = m \frac{\pi^2}{W_s^2}\,\frac{(N_s+1)(2N_s+1)}{6}.
\end{equation}

We now substitute the previous equation in Eq.\,\ref{FIM_QPD} to obtain the FIM corresponding to sparse QPD:

\begin{equation}
    I_{\mathrm{SQPD}}=m M \,p_0^2\,\bar{F}\,\frac{\pi^2}{W_s^2}\,\frac{(N_s+1)(2N_s+1)}{6}.
\end{equation}

Analogously to the previous case we can write the mean squared error bound as

\begin{equation}
\label{mse_sqpd}
    \epsilon^2 \ge\ I_{SQPD}^{-1} \ge\ \frac{3\,W_s^2}{m\pi^2M \,p_0\,(N_s+1)(2N_s+1)}.
\end{equation}

In the same way, the expected total runtime is
\begin{equation}
    \mathbb E[T_{runtime}]=2M\sum_{j=1}^{m}\mathbb E\bigl[t_{k_j}\bigr]=2mM\langle t\rangle,\qquad
    \langle t\rangle=\frac{1}{N_s}\sum_{k=1}^{N_s}t_k=\frac{\pi}{W_s}\,\frac{N_s+1}{2},
\end{equation}
i.e.\ $\mathbb E[T_{runtime}]=mM\,\frac{\pi}{W_s}\,(N_s+1)$.

We conclude this section computing the asymptotic scaling for sparse QPD. We take again the limit $N_s\to\infty$ at fixed $W_s$, with $N_s=W_sT_{max}/\pi$. Then $\langle t\rangle\simeq T_{max}/2$ and $\langle t^2\rangle=\frac{\pi^2}{W_s^2}\frac{(N_s+1)(2N_s+1)}{6}\simeq T_{max}^2/3$, so that

\begin{equation}
\label{eq:sqpd_asym}
    \epsilon^2\ \gtrsim\ \frac{3}{m\,M\,p_0^2\,\bar F\,T_{max}^2},\qquad
    \mathbb E[T_{runtime}]\simeq m\,M\,T_{max}.
\end{equation}

The number of grid points needed to complete the signal grows only logarithmically with the grid size, $m=c\,L^a$ with $L=\ln N_s$, where $c$ and $a$ are set by the recovery guarantee of the completion. At fixed $M$, Eq.\,\eqref{eq:sqpd_asym} then gives
\begin{equation}
    \epsilon\ \propto\ T_{max}^{-1}L^{-a/2}\ \propto\ T_{runtime}^{-1}L^{a/2}.
\end{equation}
For a target error $\epsilon$ this requires
\begin{equation}
    T_{max}\simeq\Bigl(\frac{3}{p_0^2\bar F\,cM}\Bigr)^{1/2}L^{-a/2}\,\epsilon^{-1},\qquad
    \mathbb E[T_{runtime}]\simeq\Bigl(\frac{3\,cM}{p_0^2\bar F}\Bigr)^{1/2}L^{a/2}\,\epsilon^{-1},
\end{equation}
i.e. Heisenberg scaling up to logarithmic factors.

We conclude by returning to the assumptions behind these bounds. First, they hold for every estimator that is unbiased for $E_0$ near its true value. Second, the bounds are necessary, not sufficient, conditions: sparse QPD reaches the scaling above only if its estimator stays within a constant factor of the bound as $N_s$ grows, and if the bias introduced by the signal completion decays faster than the statistical error.

%\bibliography{bib.bib}

\end{document}